\documentclass[journal]{IEEEtran}
 \usepackage{amsmath,amssymb}
 \usepackage{subfigure}
 \usepackage{graphicx,graphics,color,psfrag}
 \usepackage{cite,balance}
 \usepackage{caption}
 \allowdisplaybreaks
 \usepackage{algorithm}
 \usepackage{accents}
 \usepackage{amsthm}
 \usepackage{bm}
 \usepackage{algorithmic}
 \usepackage[english]{babel}
 \usepackage{multirow}
 \usepackage{enumerate}
 \usepackage{cases}
 \usepackage{stfloats}
 \usepackage{dsfont}
 \usepackage{color,soul}
 \usepackage{amsfonts}
 \usepackage{cite,graphicx,amsmath,amssymb}
 \usepackage{subfigure}
 \usepackage{fancyhdr}
 \usepackage{hhline}
 \usepackage{graphicx,graphics}
 \usepackage{array,color}
 \usepackage{amsmath}
 \usepackage{booktabs}

\newtheorem{lemma}{Lemma}
\newtheorem{corollary}{Corollary}

\newtheorem{proposition}{Proposition}

\newtheorem{remark}{\bf Remark}
\def\phi{\varphi}

\def\({\left(}
\def\){\right)}

\def\b0{{\mathbf{0}}}

\newcommand{\nn}{\nonumber}

\begin{document}

\setlength{\topskip}{-3pt}

\title{\huge Wirelessly Powered Crowd Sensing: Joint Power Transfer, Sensing, Compression, and Transmission}
\author{Xiaoyang Li, Changsheng You, Sergey Andreev, Yi Gong, and Kaibin Huang
\thanks{
This work was supported in part by Hong Kong Research Grants Council under the Grants 17209917 and 17259416, and Natural Science Foundation of Guangdong Province under Grant 2015A030313844. X. Li, C. You, and K. Huang are with the Dept. of EEE at The University of Hong Kong, Hong Kong (e-mail:  lixy@eee.hku.hk, csyou@eee.hku.hk; huangkb@eee.hku.hk). X. Li is also with the Dept. of EEE at Southern University of Science and Technology, Shenzhen, China. S. Andreev is with the Laboratory of Electronics and Communications Engineering, Tampere University of Technology, Finland (e-mail: sergey.andreev@tut.fi). Y. Gong is with the Dept. of EEE at Southern University of Science and Technology, Shenzhen, China (e-mail: gongy@sustc.edu.cn).} 
}
\maketitle

\begin{abstract}
Leveraging massive numbers of sensors in user equipment as well as opportunistic human mobility, \emph{mobile crowd sensing} (MCS) has emerged as a powerful paradigm, where prolonging battery life of constrained devices and motivating human involvement are two key design challenges. To address these, we envision a novel framework, named \emph{wirelessly powered crowd sensing} (WPCS), which integrates MCS with \emph{wireless power transfer} (WPT) for supplying the involved devices with extra energy and thus facilitating user incentivization. This paper considers a multiuser WPCS system where an \emph{access point} (AP) transfers energy to multiple \emph{mobile sensors} (MSs), each of which performs data sensing, compression, and transmission. Assuming lossless (data) compression, an optimization problem is formulated to simultaneously maximize data utility and minimize energy consumption at the operator side, by jointly controlling wireless-power allocation at the AP as well as sensing-data sizes, compression ratios, and sensor-transmission durations at the MSs. Given fixed compression ratios, the proposed optimal power allocation policy has the \emph{threshold}-based structure with respect to a defined \emph{crowd-sensing priority} function for each MS depending on both the operator configuration and the MS information. Further, for fixed sensing-data sizes, the optimal compression policy suggests that compression can reduce the total energy consumption at each MS only if the sensing-data size is sufficiently large. Our solution is also extended to the case of lossy compression, while extensive simulations are offered to confirm the efficiency of the contributed mechanisms.
\end{abstract}

\section{Introduction}
The unprecedented growth of the \emph{Internet of Things} (IoT) applications for \emph{Smart Cities} fuels the deployment of billions of wireless IoT devices that help automate a wide range of services, such as temperature measurement, pollution assessment, traffic monitoring, and public-safety surveillance. However, traditional \emph{wireless sensor networking} (WSN) solutions are limited in their coverage and scalability, as well as suffer from high maintenance costs~\cite{akyildiz2002wireless}. Recently, leveraging massive numbers of sensors in user handheld and wearable equipment led to the emergence of \emph{mobile crowd sensing} (MCS), which is becoming a new paradigm that involves humans as part of the sensing infrastructure~\cite{ganti2011mobile}. The key MCS challenges are prolonging device battery life and facilitating human engagement. These can be tackled by leveraging \emph{wireless power transfer} (WPT) techniques as a user incentive for charging \emph{mobile sensors} (MSs) within our proposed framework of \emph{wirelessly powered crowd sensing} (WPCS)~\cite{galinina2016wirelessly}. 

\begin{figure}[!t]
  \centering
  \includegraphics[scale=0.43]{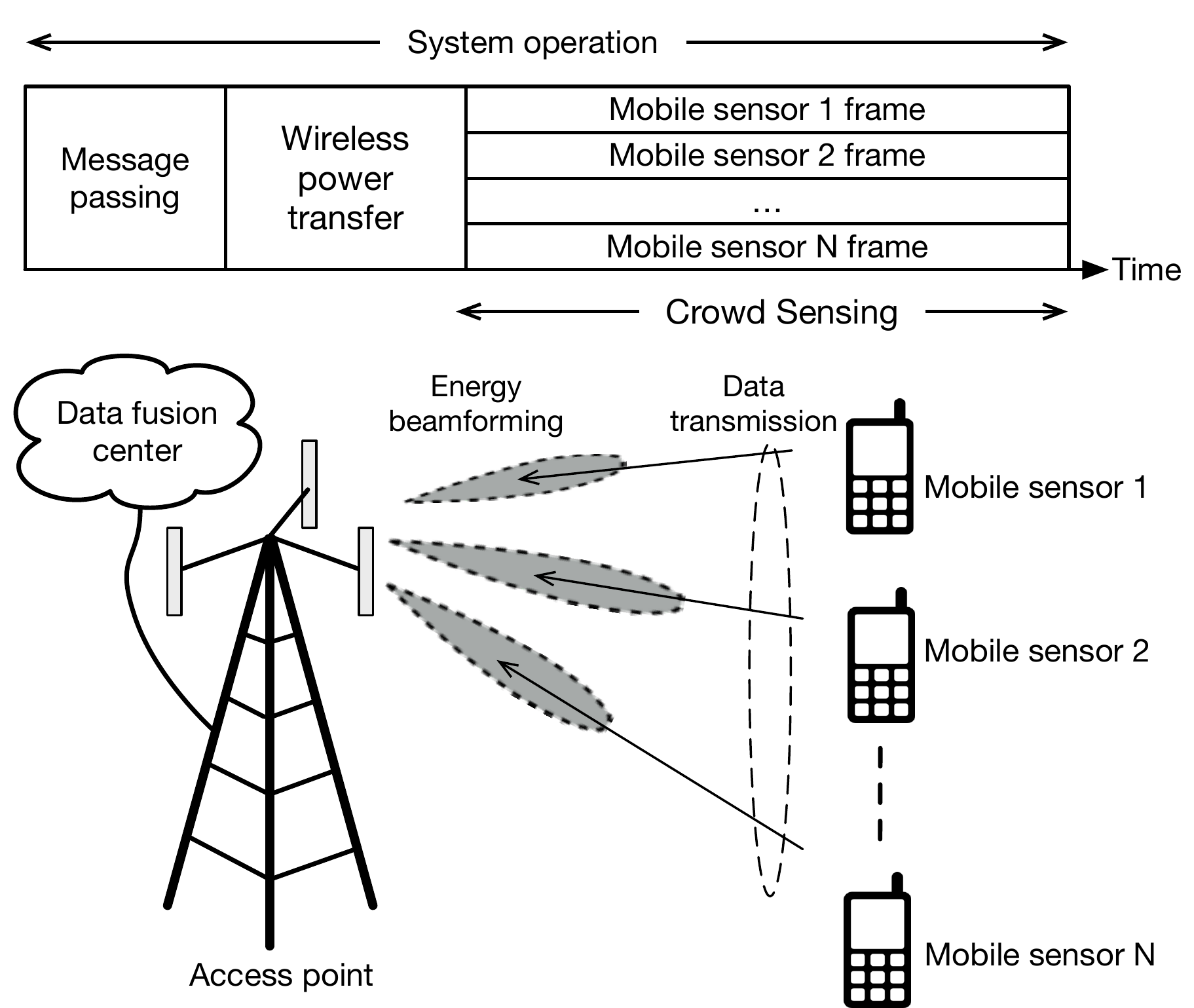}
  \caption{Our considered multiuser WPCS system.}
  \label{FigSys}
\end{figure}

\newpage 
In this work, we consider a multiuser WPCS system in Fig.~\ref{FigSys}, which comprises an operator-deployed \emph{access point} (AP) that transfers energy to multiple MSs. To optimize the operator's reward, we design a set of efficient control policies for maximizing data utility and minimizing energy consumption simultaneously.

\subsection{Mobile Crowd Sensing}
The consideration of MCS for IoT in Smart Cities has motivated research on advanced technologies that go beyond conventional WSNs, such as personal location-based systems, trusted sensing platforms, and people-centric sensing applications~\cite{ganti2011mobile}. A key MCS design issue is \emph{how to involve people into collaborate sensing}, as users may not be willing to contribute if the associated energy costs are high. There are two approaches to address this issue, known as \emph{opportunistic} and \emph{participatory} sensing~\cite{ma2014opportunities,petrov2017vehicle}. The former assumes that applications may run in the background and opportunistically collect data without user attention. The latter requires users to consciously participate in sensing activities and timely transmit the operator-requested data. However, this may consume significant device resources (i.e., battery and computing power), and it is essential to design incentivization mechanisms that compensate for the costs of participation~\cite{zhang2016incentives,feng2014trac,duan2012incentive}. 

Specifically, inspired by the location-based games, \emph{entertainment as an incentive} can enrich the experience of participants who collect e.g., geographic data on user preferences and interests. \emph{Service as an incentive} is another widely-recognized mechanism. A typical example here is traffic monitoring, wherein users provide real-time traffic information (e.g., on road congestion levels) and in return receive timely services, such as accurate route planning. Finally, \emph{money as an incentive} is an intuitive method that interprets sensed data as goods for gaining profit. Accordingly, an auction scheme for MCS named reverse auction was proposed in~\cite{feng2014trac}, where the sellers (i.e., participants) submit their bidding prices to the buyer (i.e., operator) to compete for the data sensing activities. Further, the operator selects a subset of MSs for crowd sensing operation and determines individual payments. 

In \cite{duan2012incentive}, the authors designed an alternative framework by utilizing Stackelberg game for monetary incentivization. The operator and the MSs negotiate in a two-stage manner to achieve the Stackelberg equilibrium: first the operator announces the payments and then each MS decides on whether to participate. Here, incentive allocation in multiuser MCS systems depends on data quality of each user, which has been investigated in~\cite{yang2017designing,alsheikh2017privacy}. In particular, the user data quality was estimated in~\cite{yang2017designing} by using unsupervised learning techniques for characterizing long-term reputation. For privacy management, inverse correlation between the service quality and the privacy level was determined in~\cite{alsheikh2017privacy} from the perspective of data analytics. Despite these substantial efforts, the issue of energy consumption at the MSs is not yet resolved. The proposed work addresses it by relying on advanced WPT techniques as well as contributes the first theoretical resource-allocation design of \emph{trading energy for data}, which we envisioned in~\cite{galinina2016wirelessly}.


\subsection{Wirelessly Powered Communication and Sensing}
Designed originally for point-to-point power transmission, the WPT technology has already comprised advanced techniques from inductive coupling and electromagnetic radiation \cite{lu2015wireless,Zeng2016Communications}. According to \cite{gh2016ambient}, the power to be wirelessly transferred is on the level of micro-watt, thus WPT is mostly applied for the ultra-dense small cell networks. Further integration of information and power transfer gave rise to the emerging field of \emph{simultaneous wireless information and power transmission} (SWIPT)~\cite{zhang2013mimo}. Multiple studies focused on applying SWIPT to a variety of communication systems and networks, including  \emph{multiple-input-multiple-output} (MIMO) communications \cite{zhang2013mimo,park2013joint}, two-way transmissions \cite{popovski2013interactive}, relaying \cite{ding2014power,ding2014wireless,zheng2017resource}, \emph{orthogonal frequency-division multiple access} (OFDMA) transmissions \cite{ng2013wireless}, and cognitive networking~\cite{lee2013opportunistic,ng2016multiobjective}, among others. Further design efforts tailored it for practical setups by accounting for channel state information~\cite{liu2015simultaneous} and quality of service~\cite{gregori2013energy}. Recently, WPT was considered for unmanned aerial vehicles~\cite{zeng2016throughput} and mobile-edge computing networks~\cite{you2016energy} for powering computation offloading in energy-constrained devices. 

Another important application of WPT is in sensing systems, where the MSs have a limited energy storage due to their compact form-factor, but are expected to perform sophisticated sensing tasks cooperatively. Powering these sensors is challenging, since frequent battery recharging/replacement is impractical if not impossible. Recent breakthroughs in WPT techniques offer a viable solution by powering MSs with energy beamforming~\cite{xie2013wireless,choi2017wireless,Xu2016Wireless,Guo2015Energy,han2017wirelessly}. Specifically, a multi-antenna \emph{wirelessly powered sensor networking} (WPSN) testbed has been developed in~\cite{choi2017wireless}, where a power beacon transfers energy to a sensor node via an adaptive energy-beamforming scheme. 

For multiple WPSN sensors, efficient design of energy beamforming is much more convoluted. In~\cite{Xu2016Wireless}, accounting for practical sensing and circuit power consumption at MSs, the authors proposed a collaborative energy-beamforming scheme for powering a cluster of sensors to transfer data by using distributed information beamforming. Cooperative sensing was further addressed in~\cite{Guo2015Energy}, where clusters of sensors forward their information by utilizing relay and SWIPT techniques. Finally, for low-complexity passive backscatter sensors, a novel architecture for large-scale WPSN systems was recently proposed in~\cite{han2017wirelessly}, where the sensors are powered by distributed power beacons. However, the existing literature concentrates on optimizing WPT efficiency to improve the sensing-and-communication throughput, but disregards the design of user incentivization and operator reward schemes. In this work, we bridge this gap by integrating WPT with MCS to facilitate human involvement into collective sensing activities, thus simultaneously maximizing data utility and minimizing energy consumption at the operator side. 

\subsection{Contributions and Organization}
In this work, by leveraging the potential of WPT for powering MSs and motivating MCS, we aim at addressing the following two key issues.

\begin{itemize}
\item The first one is \emph{how to incentivize and select MSs} for crowd-sensing operation. This includes designing a new incentivization approach that can trade energy for data depending on the data utility at different MSs. An optimized sensor selection policy needs to be developed at the operator to select a set of sensors to participate in the MCS for the maximum reward.

\item The second one is \emph{how to jointly control wireless-power allocation, data sensing, compression, and transmission} at the selected MSs. This requires the operator to allocate wireless power for improved energy supply of each MS, while the MSs should employ energy-efficient sensing, compression, and transmission policies to minimize the energy consumption.
\end{itemize}

To account for these two considerations, this paper addresses a multiuser WPCS system controlled by an operator and comprising a multiple-antenna AP that transfers energy to multiple single-antenna MSs. Each MS accumulates part of such energy as a reward and utilizes the rest to conduct the MCS tasks, including sensing, data compression, and transmission of compressed data to the AP. We consider both lossless and lossy (data) compression as well as design a joint control policy for maximizing the data utility and minimizing the energy consumption at the AP simultaneously. The main contributions of this work are summarized as follows.

\begin{itemize}
\item \emph{Problem Formulation and Iterative Solution:} For lossless compression, we formulate an optimization problem for simultaneously maximizing data utility and minimizing energy consumption at the operator. An iterative solution that reaches the local optimum is then proposed for a joint optimization of power allocation, sensing, compression, and transmission.

\item \emph{Joint Optimization of Power Allocation, Sensing, and Transmission}: Given fixed compression ratios, the considered non-convex problem is reduced to a convex formulation. We first derive a semi-closed form expression for the optimal sensor-transmission duration. The results are used for deriving the optimal wireless-power allocation and sensing-data sizes. 

\item \emph{Joint Optimization of Compression and Transmission}: Given fixed sensing-data sizes, the sensor-transmission durations and compression ratios are optimized for minimizing energy consumption at the operator. The derived optimal policy suggests that the MSs should perform data compression only if the sensing-data size exceeds a certain threshold. 

\item \emph{Optimal Control for Lossy Compression}: The proposed solution is further extended to lossy compression, while the optimization problem is modified to account for the respective data utility degradation. Given fixed compression ratios, the corresponding approach is similar to the lossless case. For the optimization of compression to preserve the data utility, we further optimize both the sensing-data sizes and the compression ratios at the MSs. 
\end{itemize}

The rest of this paper is organized as follows. The system model is introduced in Section II. Section III studies efficient joint control in the WPCS system based on lossless compression. In Section IV, this approach is extended for the system with lossy compression. Simulation results are offered in Section V, followed by the conclusions in Section VI.


\section{System Model}\label{Sec:Sys}

Consider a multiuser WPCS system shown in Fig.~\ref{FigSys} and comprising multiple single-antenna MSs together with one multi-antenna AP. The model of system operation is described in below sub-sections. 

\subsection{WPCS Operation}

We focus on a particular fixed time window for crowd-sensing, which is divided into three phases: \emph{message passing}, \emph{WPT}, and \emph{crowd sensing} (see Fig.~\ref{FigSys}). 
 
Consider the \emph{message-passing phase}. Each sensor feeds back to the AP its parameters including the effective channel power gain, sensing, and compression power. Given the knowledge of all sensor parameters, the AP jointly optimizes the power allocation for WPT and sensor operations (namely, data sensing, compression, and transmission). The objective is to simultaneously maximize the data utility and minimize the energy consumption (see the performance metrics in the sequel). Subsequently, the AP informs individual sensors on their allocated power, optimal compression ratios, sensing-data sizes, and time partitions for sensing, compression, and transmission. The power consumption at the AP for transmitting these control parameters is negligible due to their small data sizes.

Further, in the \emph{WPT phase}, the AP employs an antenna array to beam energy to the MSs as an incentive for sensing (see Fig.~\ref{FigSys}). Upon harvesting energy, each MS stores a part of it as a reward and applies the rest to operate the sensing tasks, including sensing, data compression, and transmission of compressed data to the AP. 
 
Finally, in the \emph{crowd-sensing phase}, the sensors simultaneously perform data sensing, compression, and transmission based on the settings determined by the AP in the message-passing phase. Parallel data collection is enabled via transmissions over orthogonal channels as allocated by the AP to the sensors based on OFDMA. The time duration for crowd sensing, denoted as $T$, is divided into three slots with \emph{adjustable} durations $t_n^{(s)}$, $t_n^{(c)}$, and $t_n$, which are used for sensing, compression, and transmission, respectively. This introduces the following constraint: 
\begin{equation}
\text{(Time constraint)} \quad t_n^{(s)} + t_n^{(c)} + t_n \leq T. \label{Eq:Time:Const}
\end{equation}

\subsection{Model of Wireless Power Transfer}

In the WPT phase with a fixed duration of $T_0$ (seconds), the AP transfers energy simultaneously to $N$ MSs by pointing $N$ radio beams to the corresponding MSs. Let $g_n$ denote the effective channel power gain between the AP and MS $n$ using energy beamforming, while $P_n$ is the transmission power of the corresponding beam. The AP transmission power is assumed to be fixed and represented by $P_0$, thus resulting in the following constraint: 
\begin{equation} \label{Eq:Power:Const}
\text{(Power constraint)} \quad \sum_{n=1}^N P_n\leq P_0.
\end{equation}
The energy transferred from the AP to MS $n$, denoted by $E_n^{(h)}(P_n)$, is  $E_n^{(h)}(P_n)=\eta g_n P_n T_0$, where the constant $0<\eta<1$ represents the energy conversion efficiency.

Each of the sensors selected to participate in the crowd sensing stores a part of received  energy as its reward. Let the reward energy at sensor  $n$, denoted by $E_n^{(r)}(\ell_n^{(s)})$, be proportional to the size of sensing data, namely,  $E_n^{(r)}(\ell_n^{(s)})=q_n^{(r)} \ell_n^{(s)}$ with $q_n^{(r)}$ being a fixed scaling factor. The remaining energy, $\left(E_n^{(h)} - E_n^{(r)}\right)$, is consumed by the sensor operations, including sensing, compression, and transmission, where energy consumption is represented respectively by $E_n^{(s)}$, $E_n^{(c)}$ and $E_n^{(t)}$ as specified  in the sequel. Based on the above discussion, each MS should satisfy the following energy constraint: 
\begin{equation}\label{Eq:Energy:Const}
\text{(Energy constraint)}~ E_n^{(r)} \!+\! E_n^{(s)} \!+\! E_n^{(c)} \!+\! E_n^{(t)} \!\leq\! \eta g_nP_n T_0.
\end{equation}

\begin{figure}[t]
  \centering
  \includegraphics[scale=0.43]{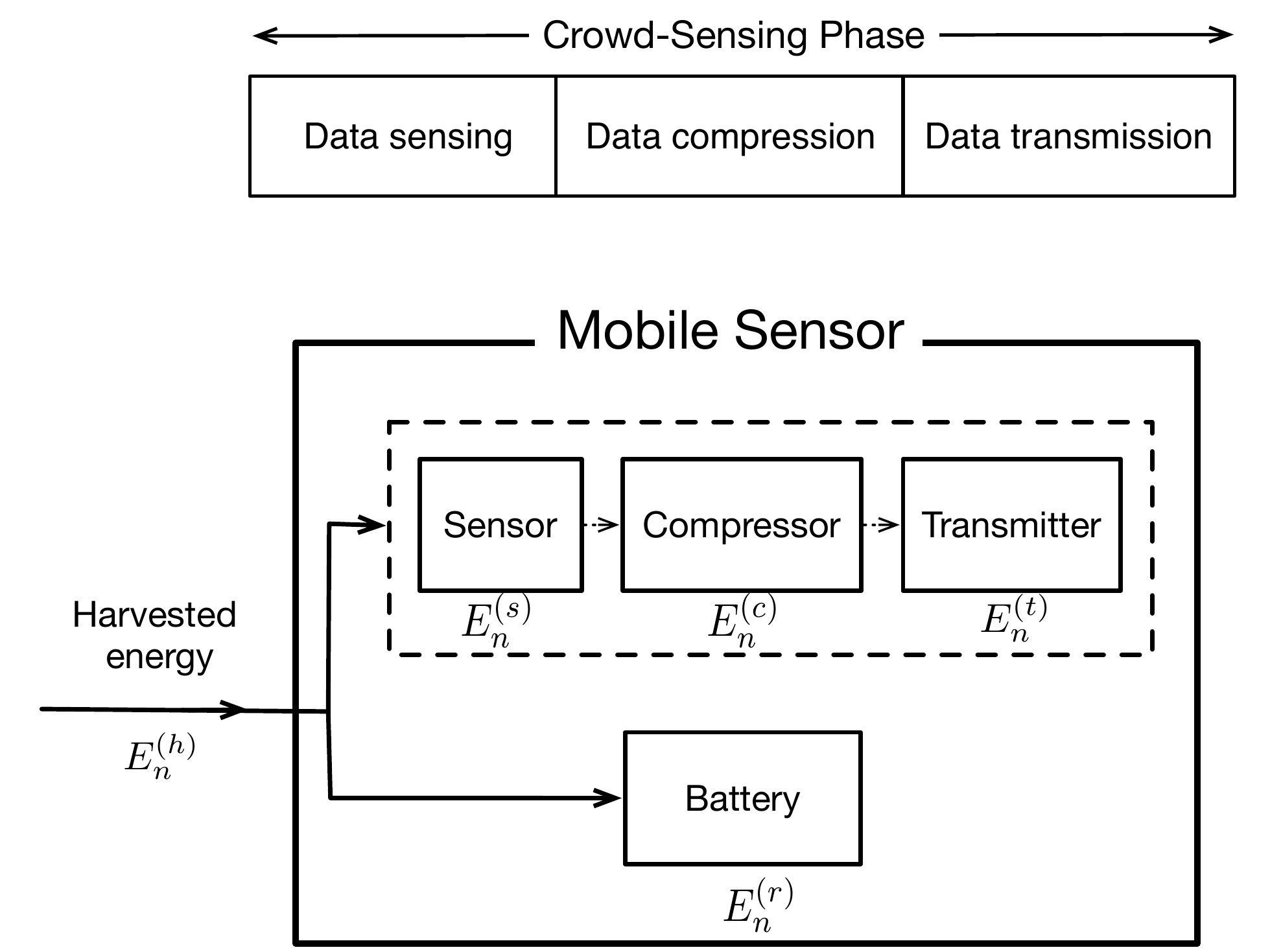}
  \caption{MS architecture, operation, and energy consumption.}
  \label{FigMob}
\end{figure}

\subsection{Mobile-Sensor Model}
\par At each MS, crowd sensing comprises three sequential operations: data sensing, data compression, and data transmission as shown in Fig.~\ref{FigMob} and modeled as follows. 

\subsubsection{Data sensing model} Consider MS $n$. Let $s_n$ denote the output data rate. Given the sensing time-duration $t_n^{(s)}$, the size of raw sensing data, denoted by $\ell_n^{(s)}$ (in bits),  is $\ell_n^{(s)}=s_n t_n^{(s)}$. Based on the data-sensing energy consumption model in~\cite{bhardwaj2001upper}, let $q_n^{(s)}$ denote the sensing energy consumption for generating $1$-bit of data. The total  energy consumption for sensing at MS $n$, denoted as  $E_n^{(s)}(\ell_n^{(s)})$, is given as $E_n^{(s)}(\ell_n^{(s)})=q_n^{(s)} \ell_n^{(s)}$.

\subsubsection{Data compression model} First, consider lossless compression of sensed data, for which the original data can be perfectly reconstructed from the compressed data. Relevant techniques include Huffman, run-length, and Lempel-Ziv encoding. It is assumed that all MSs deploy the same lossless compression method with the maximum \emph{compression ratio} denoted as $R_{\rm{max}}$, namely, the maximum ratio between the sizes of raw and compressed data. Let $R_n\in[1, R_{\rm{max}}]$ denote the compression ratio chosen by MS $n$. Then, the size of compressed data $\ell_n$ is given as  $\ell_n=\ell_n^{(s)}/R_n$. Note that in the current literature, there is still a lack of well-established models for the computation complexity of data compression. However, a tractable model can be inferred from the existing measurement results for Zlib compression \cite{Arjancompression}. Specifically, the required CPU cycles for compressing $1$-bit of data can be approximated as an exponential function of the compression ratio $R_n$ as: 
\begin{equation} \label{Eq:Comp:Complexity}
\text{(Compression complexity)} \quad C(R_n,\epsilon)=e^{\epsilon R_n}-e^{\epsilon},
\end{equation}
where $\epsilon$ is a positive constant depending on the compression method. At  $R_n=1$,  $C(R_n,\epsilon)=0$, which corresponds to no  compression.  Let $f_n$ denote the fixed CPU-cycle frequency at MS $n$ and $t_n^{(c)}$ is the compression time duration. Then,  $t_n^{(c)}=(\ell_n^{(s)} C(R_n,\epsilon))/f_n$. Following the practical model in \cite{chandrakasan1992low}, each CPU cycle consumes the energy of $q_n^{(c)}=\alpha f_n^2$, where $\alpha$ is a constant determined by the circuits. The energy consumption for data compression at MS $n$, denoted by $E_n^{(c)}(\ell_n^{(s)}, R_n)$, is given as $E_n^{(c)}(\ell_n^{(s)}, R_n)=q_n^{(c)} \ell_n^{(s)} C(R_n,\epsilon)$ with $C(R_n,\epsilon)$ in \eqref{Eq:Comp:Complexity}. It follows that 
\begin{equation} \label{Eq:Comp:Energy}
\text{(Compression energy)}~E_n^{(c)}(\ell_n^{(s)}, R_n)\!=\!q_n^{(c)}\ell_n^{(s)}(e^{\epsilon R_n}\!-\!e^{\epsilon}).\!
\end{equation}

Further, consider lossy compression that achieves a higher compression ratio than its lossless counterpart at the cost of some information loss. Typical  techniques involve data truncation after a transform (e.g., discrete cosine transform) or discarding the data immaterial for user perception. Our model of compression energy consumption follows that in \eqref{Eq:Comp:Energy} with the parameters $\{R_n, R_{\rm{max}}, \epsilon\}$ replaced by their counterparts for lossy compression $\{\check{R}_n, \check{R}_{\rm{max}}, \check{\epsilon}\}$. The  information loss can be measured with a \emph{quality factor}, denoted by $b_n\in(0,1]$ and defined as the equivalent size of raw data for $1$-bit of compressed data in terms of utility. The quality factor, however, has not been well-modeled in the existing literature. To address it, one tractable model can be inferred from  the simulation results in \cite{van1996perceptual}, given by $b_n=1/\sqrt{\check{R}_n}$. Specifically, $b_n \to 0$ when $\check{R}_n \to \infty$, and $b_n \to 1$ when $\check{R}_n \to 1$, which corresponds to no compression.

\subsubsection{Data transmission model} Each selected MS transmits its compressed data to the AP. Let $P_n^{(t)}$ denote the transmission power and $t_n$ is the transmission time duration. Assuming channel reciprocity, the achievable transmission rate (in bits/s), denoted by $v_n$, can be given as $v_n=\ell_n/t_n=B\log_2(1+\frac{g_n P_n^{(t)}}{N_0})$, where $B$ is the bandwidth and $N_0$ is the variance of complex-white-Gaussian noise. Hence, the transmission energy consumption denoted by $E_n^{(t)}(\ell_n)$ follows: $E_n^{(t)}(\ell_n)=P_n^{(t)} t_n=\dfrac{t_n}{g_n}f(\ell_n/t_n)$, where the function $f(x)$ is defined as $f(x)=N_0(2^{\frac{x}{B}}-1)$.

\subsubsection{Time and energy constraints (revisited)}

Finally, based on the above models, the time constraint in \eqref{Eq:Time:Const} and energy constraint in \eqref{Eq:Energy:Const} can be rewritten as 
\begin{align}
&\frac{\ell_n^{(s)}}{s_n}+\frac{\ell_n^{(s)} C(R_n,\epsilon)}{f_n}+t_n\le T,\label{Eq:Time:Const:a}\\
&\left[q_n^{(r)}\!+\!q_n^{(s)}\!+\!q_n^{(c)} C(R_n,\epsilon)\right] \ell_n^{(s)}\!+\!\dfrac{t_n}{g_n} f\left(\frac{\ell_n^{(s)}}{t_n R_n}\right) \!\le\! \eta g_n P_n T_0.\label{Eq:Energy:Const:a}
\end{align}

An important observation from the time constraint perspective is that given the sensing rate $s_n$ and the CPU-cycle frequency $f_n$, the partitioning of crowd-sensing time of sensor $n$ for sensing, compression, and transmission can be determined by the \emph{sensing-data size $\ell_n$}, \emph{compression ratio $R_n$}, and \emph{transmission time $t_n$} to be optimized in the sequel. 

\subsection{Performance Metrics}
Recall that the \emph{operator's reward}, which becomes our performance measure, refers to  the difference between the sum weighted data utility and the weighted energy cost. Its mathematical definition is provided as follows.  Following a commonly used model (see e.g., \cite{yang2012crowdsourcing}), the utility of $\ell_n$-bit data delivered by sensor $n$ is tackled by the logarithmic function $a_n \log\left(1+b_n \ell_n^{(s)}\right)$, where $b_n$ is the said information loss due to compression and $a_n$ is a weight factor depending on the type of data. Note that the  logarithmic function is chosen here to model the diminishing return as the data size increases. Hence, the sum data utility for the operator, denoted by $U(\boldsymbol{\ell}^{(s)})$, is 
\begin{equation} \label{Eq:Utility}
U(\boldsymbol{\ell}^{(s)})=\sum_{n=1}^{N}{a_n}\log(1+b_n \ell_n^{(s)}).
\end{equation}
 
Then, the operator's reward, denoted by $R(\boldsymbol{\ell}^{(s)},\boldsymbol{P})$, can be modeled as
\begin{align} \label{Eq:Reward}
R(\boldsymbol{\ell}^{(s)},\boldsymbol{P})=\sum_{n=1}^{N} a_n \log(1+b_n \ell_n^{(s)})-c\sum_{n=1}^N P_n T_0,
\end{align}
where $c$ denotes the price of unit energy with respect to that of unit data utility.


\section{Joint Power Allocation, Sensing, Compression, and Transmission}
In this section, we formulate and solve a non-convex problem of jointly optimizing the power allocation, sensing, lossless compression, and transmission by deriving the local-optimal policy as discussed in Section II. This yields important guidelines for operating the proposed WPCS system. The results are extended to the case of lossy compression in the following section. 
 
\subsection{Problem Formulation} 
The specific design problem here is to jointly  optimize the AP power allocation for WPT to sensors, $\{P_n\}$, the sizes of sensing data, $\{\ell_n^{(s)}\}$, the data compression ratios, $\{R_n\}$, and the partitioning of crowd-sensing time for sensing and compression, determined by $\{t_n\}$ together with $\{\ell_n^{(s)}\}$ and $\{R_n\}$. The objective is to maximize the operator's reward in \eqref{Eq:Reward} under the power constraint in \eqref{Eq:Power:Const}, time constraint in \eqref{Eq:Time:Const:a}, and energy constraint in \eqref{Eq:Energy:Const:a}. For lossless compression, the quality factor $b_n=1$. Mathematically, the optimization problem at hand can be formulated as follows:
\begin{equation*}
\begin{aligned}
\max_{\substack{P_n\ge0, \ell_n^{(s)} \ge0,\\ R_n\in [1, R_{\rm{max}}], t_n\ge0}} \quad 
&\sum_{n=1}^{N} a_n \log(1+\ell_n^{(s)})-c\sum_{n=1}^{N} P_n T_0\\ 
\text{s.t.} \qquad
&\sum_{n=1}^{N} P_n \le P_0,\\
\textbf{(P1)}\qquad \qquad &\frac{\ell_n^{(s)}}{s_n}+\frac{\ell_n^{(s)} C(R_n,\epsilon)}{f_n}+t_n\le T,\qquad~~ \forall~n, \\
&[q_n^{(r)}+q_n^{(s)}+q_n^{(c)}C(R_n,\epsilon)]\ell_n^{(s)}\\
&\qquad +\dfrac{t_n}{g_n} f\left(\frac{\ell_n^{(s)}}{t_n R_n}\right)\le \eta g_n P_n T_0,\quad \forall~n.
\end{aligned}
\end{equation*}

\begin{figure}[t!]
  \centering
  \includegraphics[scale=0.38]{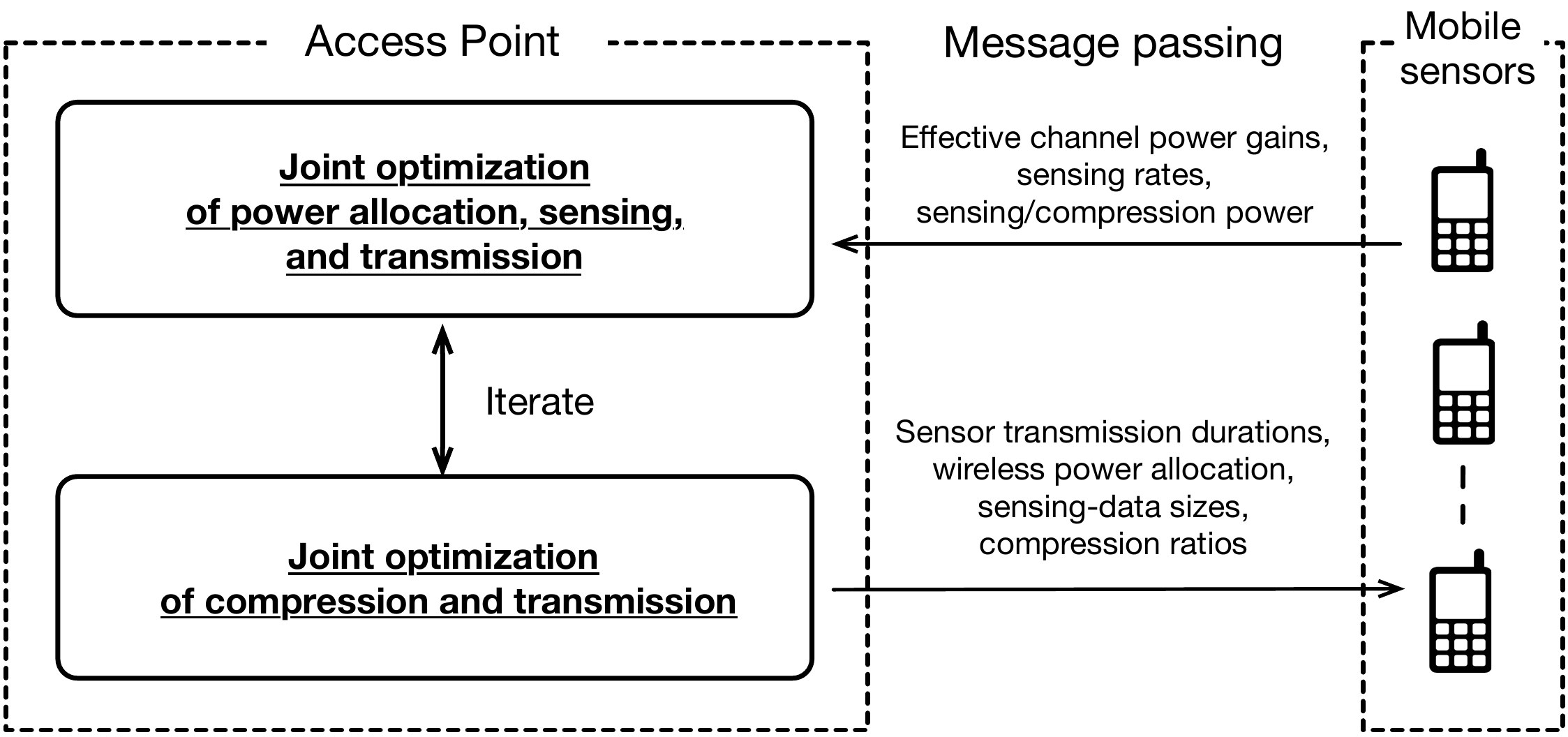}
  \caption{Implementation of our iterative  solution  for optimal WPCS.}
  \label{FigNego}
\end{figure}

\subsection{Iterative Solution Approach} \label{Sec:Iterative}
The convexity of problem P1 is characterized  as follows.
\begin{lemma}\label{Lem:ConvP1}\emph{Problem P1 is a \emph{non-convex} optimization problem.}
\end{lemma}
This lemma can be easily proved, since the optimization variables, $\left\{\ell_n^{(s)}, R_n, t_n\right\}$, are coupled at the constraints in forms of $\ell_n^{(s)} C(R_n,\epsilon)$ and $\dfrac{t_n}{g_n}f\left(\ell_n^{(s)}/(t_n R_n)\right)$. To establish the optimal policy for attaining a \emph{local} maximum and  characterize its structure, we propose an \emph{iterative}  solution approach as demonstrated in Fig.~\ref{FigNego}. Specifically, the approach iterates between solving two  sub-problems, P1-A and P1-B as given below, for optimizing two subsets of parameters, namely, $\{P_n, \ell_n^{(s)}, t_n\}$ and $\{R_n, t_n\}$, respectively. The reason for such separation  of parametric optimization is in that it creates two convex sub-problems as shown in the sequel. It is necessary to include the transmission durations $\{t_n\}$ as variables in both sub-problems, since they are affected by all other parameters.
 
First, with the compression ratios $\{R_n\}$ fixed, problem P1 reduces to the following problem P1-A, named \emph{joint optimization of power allocation, sensing, and transmission}:  
\begin{equation*}
\begin{aligned}
\max_{\substack{P_n\ge0, \ell_n^{(s)} \ge0,\\ t_n\ge0}} \quad 
&\sum_{n=1}^{N}a_n\log(1+\ell_n^{(s)})-c\sum_{n=1}^{N} P_n T_0\\ 
\text{s.t.} \qquad
&\sum_{n=1}^{N} P_n \le P_0,\\
\textbf{(P1-A)}\qquad 
&\beta_n \ell_n^{(s)}+t_n\le T,\qquad \qquad \qquad \qquad \qquad \forall~n,\\
&\alpha_n \ell_n^{(s)}+\dfrac{t_n}{g_n} f\left(\frac{\ell_n^{(s)}}{R_n t_n}\right)\le \eta g_n P_n T_0, ~~~~~ \forall~n,
\end{aligned}
\end{equation*}
where $\alpha_n=q_n^{(r)}+q_n^{(s)} +q_n^{(c)} C(R_n,\epsilon)$ and $\beta_n=\frac{1}{s_n}+\frac{C(R_n,\epsilon)}{f_n}$. Next, given  the parameters $\{P_n\}\cup\{\ell_n^{(s)}\}\cup\{t_n\}$, the sensing-data sizes $\boldsymbol{\ell}^{(s)}$ are fixed and thus  problem P1 is reduced to minimizing the sum energy consumption: $\sum_{n=1}^{N} P_n T_0$. Given this criterion, the energy constraints in \eqref{Eq:Energy:Const:a} are active at the optimal point: 
\begin{equation}
\left[q_n^{(r)}\!+\!q_n^{(s)}\!+\!q_n^{(c)} C(R_n,\epsilon)\right]\ell_n^{(s)}\!+\!\dfrac{t_n}{g_n} f\left(\frac{\ell_n^{(s)}}{t_n R_n}\right)= \eta g_n P_n T_0.
\end{equation}

After substituting this, problem P1-B, named \emph{joint optimization of compression and transmission}, can be produced from problem P1 as:
\begin{equation*}
\begin{aligned}
\min_{\substack{R_n\in [1, R_{\rm{max}}],\\ t_n\ge 0}}~
& [q_n^{(r)}\!+\!q_n^{(s)}\!+\!q_n^{(c)} C(R_n,\epsilon)]\ell_n^{(s)}\!+\!\dfrac{t_n}{g_n} f\left(\frac{\ell_n^{(s)} }{t_n R_n}\right)\\
\textbf{(P1-B)} \quad \text{s.t.} \quad 
&\frac{\ell_n^{(s)}}{s_n}+\frac{\ell_n^{(s)} C(R_n,\epsilon)}{f_n}+t_n\le T, \quad \quad \quad ~\forall~n.
\end{aligned}
\end{equation*}  
By iteratively solving these two problems, which are proved to be convex in the sequel, the solution is guaranteed to \emph{converge} and reach a \emph{local} maximum of the operator's reward  function in formulation P1. 

\subsection{Joint Optimization of Power Allocation, Sensing, and Transmission}\label{Sec:JointFixedCompress}
To solve problem P1-A, we first prove that it can be reduced to the problem of transmission-duration  optimization. Then, the solution of the reduced problem is used to characterize the optimal wireless power allocation and sensing-data sizes. 

\subsubsection{Optimal Sensor Transmission} 
Problem P1-A is solved as follows. First, observe that problem P1-A is always feasible, since $\{P_n=0, \ell_n^{(s)}=0, R_n=1\}$ for each $n$ is a feasible solution. Next, some useful necessary conditions for the optimal solution are provided as follows. 

\begin{lemma}\label{Lem:NecCond}
\emph{As a result of solving problem P1-A, the optimal power allocation $\{P_n^*\}$, sensor transmission durations $\{t_n^*\}$,  and sensing-data sizes $\{(\ell_n^{(s)})^*\}$ satisfy the following:
\begin{align*}
&\alpha_n \left(\ell_n^{(s)}\right)^*+\dfrac{t_n^*}{g_n} f\left(\frac{(\ell_n^{(s)})^*}{R_n t_n^*}\right)=\eta g_n P_n^* T_0, &&~ \forall n, \\
& \beta_n \left(\ell_n^{(s)}\right)^*+t_n^*= T, &&~ \forall n.\label{NecCond}
\end{align*}
}
\end{lemma}
Lemma~\ref{Lem:NecCond} is proved in Appendix~\ref{App:NecCond}. The equalities arise from the fact that to maximize the operator's reward, each MS should \emph{fully} utilize the transferred energy and the crowd-sensing time. Based on these equalities, $\{P_n^*\}$ and $\{(\ell_n^{(s)})^*\}$ can be written as functions of $\{t_n^*\}$. Consequently, problem P1-A can be equivalently transformed into problem P2 for sensor-duration optimization having only $\{t_n\}$ as its variables: 
\begin{equation*}
\begin{aligned}
\max_{t_n\ge0}\quad 
&\sum_{n=1}^{N}a_n\log\left(1+\frac{T-t_n}{\beta_n}\right)\\
&\qquad -c\sum_{n=1}^N \left[\frac{\alpha_n(T-t_n)}{\eta \beta_n g_n}+\dfrac{t_n}{\eta g_n^2} f\left(\frac{T-t_n}{{R_n}{\beta_n}{t_n}}\right)\right]\\ 
\textbf{(P2)}~~\text{s.t.} \quad 
&\sum_{n=1}^N \left[\frac{\alpha_n(T-t_n)}{\eta \beta_n g_n}+\dfrac{t_n}{\eta g_n^2} f\left(\frac{T-t_n}{R_n \beta_n t_n}\right)\right]\le P_0 T_0, \\
&0\le t_n\le T, \qquad\qquad\qquad\qquad\qquad\qquad~~~~ \forall~n.
\end{aligned}
\end{equation*}

The convexity of problem P2 is established in the following lemma as proved in Appendix~\ref{App:conv}.
\begin{lemma}\label{Lem:conv}
\emph{Problem P2 is a convex optimization problem.}
\end{lemma}
As a result, problem P2 can be solved by the Lagrange method. The corresponding partial Lagrange function is 
\begin{align}
L(t_n,\lambda)=
&\sum_{n=1}^{N} (\lambda\!+\!c)\left[\frac{\alpha_n(T\!-\!t_n)}{\eta \beta_n g_n}\!+\!\dfrac{t_n}{\eta g_n^2} f\left(\frac{T\!-\!t_n}{R_n \beta_n t_n}\right)\right] \nonumber \\
&\quad\qquad-a_n\log\left(1\!+\!\frac{T\!-\!t_n}{\beta_n}\right)\!-\!\lambda P_0 T_0.
\end{align}
Define a function $y(x)$ as $y(x)=f(x)-(x+\frac{1}{R_n \beta_n})f'(x)$ and $t_n^*$ as the optimal solution for solving problem P2. Then, applying \emph{Karush-Kuhn-Tucker} (KKT) conditions leads to the following necessary and sufficient conditions:
\begin{subequations}
\begin{align}
&\frac{\partial L}{\partial t_n^*}=\frac{a_n}{\beta_n\!+\!(T\!-\!t_n^*)}\!+\!\frac{(\lambda^*\!+\!c_n)}{\eta g_n}\left[\frac{1}{g_n}y\left(\frac{T-t_n^*}{R_n \beta_n t_n^*}\right)\!-\!\frac{\alpha_n}{\beta_n}\right]\nonumber\\
&\qquad\begin{cases}
>0,&t_n^*=0,\\
=0,&0<t_n^*<T,\\
<0,&t_n^*=T,
\end{cases}\label{Eq:KKT1}\\
&\lambda^*\left(\sum_{n=1}^{N}\left[\frac{t_n^*}{\eta g_n^2}f\left(\frac{T\!-\!t_n^*}{R_n \beta_n t_n^*}\right)\!+\!\frac{\alpha_n(T\!-\!t_n^*)}{\eta \beta_n g_n}\right]\!-\!P_0 T_0\right)\!=\!0. \label{Eq:KKT2}
\end{align}
\end{subequations}
Combing these conditions yields the optimal sensor-transmission policy given as follows.

\begin{proposition}[Optimal Sensor Transmission]\label{Pro:LosslessSensorT}\emph{Given fixed compression ratios, for each MS selected for crowd sensing with $(0<t_n^*\le T)$, the optimal transmission duration $t_n^*$ satisfies:
\begin{align}\label{Eq:LosslessOpt}
&\frac{(\lambda^*+c) N_0}{\eta g_n^2}\left[\left(1-\frac{T\ln2}{B R_n \beta_n t_n^*}\right)2^\frac{T-t_n^*}{B R_n \beta_n t_n^*}-1\right] \nn\\
&\qquad\qquad\qquad+\frac{a_n}{\beta_n+(T-t_n^*)}-\frac{(\lambda^*+c)\alpha_n}{\eta g_n \beta_n}=0,
\end{align}
where $\lambda^*$ satisfies the condition in \eqref{Eq:KKT2}.
}
\end{proposition}

One can observe from Proposition~\ref{Pro:LosslessSensorT} that the optimal sensor-transmission duration $t^*_n$ has no closed form. However, the value of $t^*_n$ can be efficiently computed by following the proposed bisection-search procedure summarized in Algorithm~\ref{Al:Transmission}. Based on Proposition~\ref{Pro:LosslessSensorT}, the dependence of the optimal sensor-transmission durations on the utility weights and channels is characterized in the following corollary.

\begin{algorithm}[t!]
  \caption{A Bisection-Search Algorithm for Computing the Optimal Sensor Transmission}
  \label{Al:Transmission}
  \begin{itemize}
\item{\textbf{Step 1} [Initialize]: Let $\lambda_{\ell}=0$, $\lambda_h=\lambda_{\rm{max}}$,  and $\xi >0$.\\
Obtain $t_{n,\ell}$ and $t_{n,h}$ based on \eqref{Eq:LosslessOpt}, calculate $E_{\ell}=\sum_{n=1}^N \left[\frac{\alpha_n(T-t_{n,\ell})}{\eta \beta_n g_n}+\frac{t_{n,\ell}}{\eta g_n^2} f\left(\frac{T-t_{n,\ell}}{R_n \beta_n t_{n,\ell}}\right)\right]$ and $E_h = \sum_{n=1}^N \left[\frac{\alpha_n(T-t_{n,h})}{\eta \beta_n g_n}+\frac{t_{n,h}}{\eta g_n^2} f\left(\frac{T-t_{n,h}}{R_n \beta_n t_{n,h}}\right)\right]$ respectively.}
\item{\textbf{Step 2}  [Bisection search]: \emph{While} $E_{\ell} \neq P_0 T_0$ and $E_h \neq P_0 T_0$, update $\lambda_{\ell}$ and $\lambda_h$ as follows.\\
(1) Define $\lambda_m\!=\!(\lambda_{\ell}\!+\!\lambda_h)/2$, compute $t_{n,m}$ and $E_m\!=\!\sum_{n=1}^N \left[\frac{\alpha_n(T-t_{n,m})}{\eta \beta_n g_n}\!+\!\frac{t_{n,m}}{\eta g_n^2} f\left(\frac{T-t_{n,m}}{R_n \beta_n t_{n,m}}\right)\right]$.\\
(2) If $E_m<P_0 T_0$, let $\lambda_h=\lambda_m$, else $\lambda_{\ell}=\lambda_m$.\\
\emph{Until} $\lambda_h-\lambda_{\ell}<\xi$. Return $t_n^*=t_{n,m}$.}
\end{itemize}
  \end{algorithm}

\begin{corollary}[Properties of Optimal Sensor Transmission]\label{Cor:StruTime}\emph{The optimal sensor-transmission durations $\{t_n^*\}$ have the following properties:  
\begin{itemize}
\item[\emph{1)}] If the effective channel power gains $\{g_n\}$ are identical but the utility weights satisfy  $a_1\ge a_2\cdots \ge a_N$, then $t_1^*\le t_2^* \cdots \le t_N^*$.
\item[\emph{2)}] If the utility weights $\{a_n\}$ are identical but the effective channel power gains satisfy $g_1\ge g_2\cdots  \ge g_n$, then $t_1^*\le t_2^* \cdots \le t_N^*$. 
\end{itemize} }
\end{corollary}

The proof is given in Appendix~\ref{App:StruTime}. These cases indicate that for a sensor with large utility weight and high effective channel power gain, it is beneficial to have a short transmission duration so that the sensing duration can be made longer to increase the amount of sensing data, for larger sum data utility.

Last, the following result for the case of small optimal transmission durations ($t_n^*\ll T$)  follows from  Proposition~\ref{Pro:LosslessSensorT} by substituting $T-t_n^*\approx T$. 
\begin{corollary}[Approximate Optimal Sensor Transmission]\label{Lem:LosslessAppro}\emph{Consider the optimal sensor-transmission duration $t_n^*$ in \eqref{Eq:LosslessOpt}. If $t_n^*\ll T$, 
\begin{equation}\label{Eq:LosslessAppro}
t_n^*\approx \frac{T\ln2}{B R_n \beta_n (\tilde{\theta}-1)}, 
\end{equation}
where $\tilde{\theta}\!=\!W_0\left[\left(\dfrac{a_n \eta g_n^2}{(c\!+\!\lambda^*) (\beta_n\!+\!T) N_0 }\!-\!\dfrac{g_n \alpha_n}{N_0 \beta_n}\!-\!1\right)e^{\frac{\ln2}{B R_n \beta_n}\!-\!1}\right]$.}
\end{corollary}

Here, one can observe that the approximate result retains the properties in Corollary~\ref{Cor:StruTime}. Moreover, this result shows that the optimal sensor-transmission duration increases with the growing sensing-and-compression power as represented by $\alpha_n$ and defined in problem P1-A.

\subsubsection{Optimal Wireless-Power Allocation}
Combining the optimal sensor-transmission policy in Proposition~\ref{Pro:LosslessSensorT} with the optimality conditions in Lemma~\ref{Lem:NecCond} yields the optimal wireless-power allocation policy formulated in the following proposition, which is proved in Appendix~\ref{App:LosslessPowerAllo}. 

\begin{proposition}[Optimal Wireless-Power Allocation]\label{Pro:LosslessPowerAllo}\emph{The optimal wireless-power allocation policy $\{P_n^*\}$ that solves  problem P1-A is given as 
\begin{equation}\label{Eq:OptPower}
P_n^*\!=\!
\begin{cases}
\dfrac{1}{\eta g_n T_0}\left[\dfrac{t_n^*}{g_n}f\left(\dfrac{T\!-\!t_n^*}{R_n \beta_n t_n^*}\right)\!+\!\dfrac{\alpha_n (T\!-\!t_n^*)}{\beta_n}\right],&\phi_n \geq \lambda^*,\\
0,&\phi_n < \lambda^*,
\end{cases}
\end{equation}
where $t_n^*$ and $\lambda^*$ are given in Proposition~\ref{Pro:LosslessSensorT}, while $\phi_n$ is named the \emph{crowd-sensing priority function} defined as
\begin{equation}\label{Eq:LosslessPrio}
\phi_n\!=\!\kappa_n-c,\\
\end{equation}
where
\vspace{-10pt}
\begin{equation}\label{Eq:Kappa}
\kappa_n \overset{\triangle}{=}\!\frac{a_n \eta g_n}{q_n^{(r)}+q_n^{(s)}+q_n^{(c)} C(R_n,\epsilon)+\frac{N_0\ln2}{g_n B R_n}}.
\end{equation}}
\end{proposition}

Proposition~\ref{Pro:LosslessPowerAllo} yields that the optimal wireless-power allocation policy has a \emph{threshold}-based structure. In other words, only the MSs with (crowd-sensing) priority functions exceeding the threshold $\lambda^*$ will be allocated wireless power or, equivalently, selected for participating in the crowd sensing operation. Specifically, both the operator's configuration (including cost weight as well as provisioned power), and MSs' individual parameters significantly affect the crowd-sensing sensor selection as discussed in the following remarks.

\begin{remark}[Effects of Operator's Configuration on Sensor Selection]\emph{First, it can be observed from \eqref{Eq:LosslessPrio} that if the cost weight $c$ is sufficiently large ($c> \max\{\kappa_n\}$), the priority function for each sensor is negative i.e., $\phi_n\le 0, \forall n$. Since the Lagrange multiplier $\lambda^*\ge0$, it follows that $\phi_n< \lambda^*, \forall n$, thus implying that no sensor is selected for crowd sensing. On the other hand,  if the  energy cost is negligible (i.e., $c=0$), each sensor has a positive priority function. In this case, provided that the operator supplies sufficient power for crowd sensing, which leads to $\lambda^*=0$ (see \eqref{Eq:KKT1}), all of the sensors will be selected to participate in the crowd sensing to increase the operator's reward. For the general  case of finite $c$ and $\lambda^*>0$, it is desirable for the operator to choose only a subset of sensors with the priorities above $\lambda^*$ for crowd sensing.}
\end{remark}

\begin{remark}[Effects of Sensor Parameters on Sensor Selection]\emph{Individual parameters of sensors also affect the sensor-selection policy by the term of $\kappa_n$ as defined in \eqref{Eq:Kappa}. To be specific, a sensor with smaller sensing-and-compression power and larger effective channel power gain has higher crowd-sensing priority and thus is more likely to be selected. Such sensors with larger effective channel power gains can not only harvest more energy, but also consume less transmission power for their data transmission. Last, to compete for crowd sensing with other sensors, a sensor can reduce its amount of reward power, $q_n^{(r)}$, so as to increase its own priority function.}
\end{remark}

Finally, the sum allocated power is affected by the cost weight as shown below.

\begin{corollary}\emph{The sum allocated power $\sum_{n=1}^N P_n^* T_0$ decreases with the cost weight $c$. In particular, $\sum_{n=1}^N P_n^* T_0=P_0 T_0$, if $c=0$.}
\end{corollary}

This corollary is aligned with intuition and suggests that the optimal wireless power allocation balances the tradeoff between obtaining data utility and energy with the price measured by $c$. 

\subsubsection{Optimal Sensing-Data Sizes}
Using a similar approach for deriving the optimal wireless power allocation in Proposition~\ref{Pro:LosslessPowerAllo}, the optimal sensing-data sizes can be derived in the following proposition by using Proposition~\ref{Pro:LosslessSensorT} and Lemma~\ref{Lem:NecCond}.
\begin{proposition}[Optimal Sensing-Data Sizes]\label{Pro:LosslessDataSize}\emph{The optimal sensing-data sizes $\{(\ell_n^{(s)})^*\}$ that solve problem P1-A are given as: 
\begin{equation}
(\ell_n^{(s)})^*=
\begin{cases}
\dfrac{T-t_n^*}{\beta_n},&\phi_n \geq \lambda^*,\\
0,&\phi_n < \lambda^*,
\end{cases}
\end{equation}
where $t_n^*$ and $\lambda^*$ are given in Proposition~\ref{Pro:LosslessSensorT}, while $\phi_n$ is  defined in \eqref{Eq:LosslessPrio}.}
\end{proposition}
The effects of parameters on the sensing-data size are characterized as follows.

\begin{corollary}[Properties of Optimal Sensing-Data Sizes]\label{Cor:DataSize}\emph{The optimal sensing-data sizes $\{(\ell_n^{(s)})^*\}$ have the following properties:  
\begin{itemize}
\item[\emph{1)}] If the effective channel power gains $\{g_n\}$ are identical but the utility weights satisfy $a_1\ge a_2\cdots \ge a_N$, then $(\ell_1^{(s)})^*\ge (\ell_2^{(s)})^* \cdots \ge (\ell_n^{(s)})^*$.
\item[\emph{2)}] If the utility weights $\{a_n\}$ are identical but the effective channel power gains satisfy $g_1\ge g_2\cdots \ge g_n$, then $(\ell_1^{(s)})^*\ge (\ell_2^{(s)})^* \cdots \ge (\ell_n^{(s)})^*$.
\end{itemize} }
\end{corollary}
This corollary reflects that it is desirable to increase the sensing-data sizes for sensors with high effective channel power gains, since they consume less transmission energy.

\subsection{Joint Optimization of Compression and Transmission}
Consider the optimization formulation in problem P1-B (see Section~\ref{Sec:Iterative}). First, the problem convexity is established in the following lemma, which can be easily proved by using the property of \emph{perspective function} \cite{Boyd2006convex}.

\begin{lemma}\label{Lem:losslessconv}
\emph{Problem P1-B is a convex optimization problem.}
\end{lemma}

Hence, problem P1-B can be solved by the Lagrange method. The Lagrange function is 
\begin{align}
\!\!\widehat{L}(R_n,t_n,\widehat{\lambda})
&=Q(R_n)\ell_n^{(s)}+\dfrac{t_n}{g_n}f\left(\frac{\ell_n^{(s)}}{t_n R_n}\right)\nn \\
&\quad+\widehat{\lambda}\left(t_n+\frac{\ell_n^{(s)}C(R_n,\epsilon)}{f_n}+\frac{\ell_n^{(s)}}{s_n}-T\right),
\end{align}
where $Q(R_n)=q_n^{(r)}+q_n^{(s)}+q_n^{(c)}C(R_n,\epsilon)$. Let $g(x)=f(x)-x f^{'}(x)$ and $t_n^*$, $R_n^*$ be the optimal solution for tackling problem P1-B. Directly applying the KKT conditions results in the following necessary and sufficient conditions:
\begin{subequations}\label{Eq:LosslessKKT}
\begin{align}
&\frac{\partial \widehat{L}}{\partial R_n^*}=\left(q_n^{(c)}+\frac{\widehat{\lambda}^*}{f_n}\right)\ell_n^{(s)} \epsilon e^{\epsilon R_n^*}-\frac{\ell_n^{(s)}}{g_n R_n^{*2}}f'\left(\frac{\ell_n^{(s)}}{t_n^* R_n^*}\right)\nonumber\\
&\qquad\begin{cases}
>0,&R_n^*=1,\\
=0,&1<R_n^*<R_{\rm{max}},\\
<0,&R_n^*=R_{\rm{max}},
\end{cases}\label{Eq:LosslessKKT1}\\
&\frac{\partial{\widehat{L}}}{\partial{t_n^*}}=\frac{1}{g_n}g\left(\frac{\ell_{n}^s}{t_n^* R_n^*}\right)+{\widehat{\lambda}}^{*}
\begin{cases}
>0, &t_n^*=0,\\
=0, &0<t_n^*<T,\\
<0, &t_n^*=T,
\end{cases}\label{Eq:LosslessKKT2}\\
&\widehat{\lambda}^*\left(t_n^*+\frac{\ell_n^{(s)}C(R_n^*,\epsilon)}{f_n}+\frac{\ell_n^{(s)}}{s_n}-T\right)=0.\label{Eq:LosslessKKT3}
\end{align}
\end{subequations}

Therefore, the optimal sensor decisions on \emph{compress-or-not} are derived as follows, which is proved in Appendix~\ref{App:CompOrNot}. 

\begin{lemma}[Compress or Not?]\label{Lem:CompOrNot}\emph{Given fixed sensing-data sizes, sensor $n$ should perform data compression (i.e., $R_n>1$) for minimizing its energy consumption, if and only if its sensing-data size satisfies: $$\ell_n^{(s)}>\frac{T}{\widehat{\theta}},$$ where
$\widehat{\theta}=\frac{\ln2}{B \left[W_0\left(-\left(\dfrac{q_n^{(c)} g_n f_n}{N_0}+1\right)e^{-\vartheta_n}\right)+\vartheta_n+1\right]}+\dfrac{1}{s_n}$, and $\vartheta_n={\frac{f_n \ln2}{B \epsilon e^{\epsilon}}+1}$.}
\end{lemma}

Lemma~\ref{Lem:CompOrNot} suggests that the optimal decision on compress-or-not has a \emph{threshold-based} structure. To be specific, data compression is preferred only if the sensing-data size exceeds a threshold that depends on the effective channel power gain and compression power among others. This is due to the fact that reducing the data sizes by compression can substantially lower the exponentially-increasing energy required for transmitting larger amounts of data.

Based on the conditions in \eqref{Eq:LosslessKKT1}-\eqref{Eq:LosslessKKT3}, the key results of this subsection are derived in the following proposition.

\begin{proposition}[Optimal Compression and Transmission]\label{Pro:LosslessDataComp}\emph{The solution of problem P1-B is as follows. 
\begin{itemize}
\item[1)] The optimal sensor-transmission durations $\{t_n^*\}$ are
\begin{equation}
t_n^*= T -\dfrac{\ell_n^{(s)}\left(e^{\epsilon R_n^*}-e^{\epsilon}\right)}{f_n} -\dfrac{\ell_n^{(s)}}{s_n}, \qquad \forall \ n,
\end{equation}
which corresponds to full utilization of the crowd-sensing duration.
\item[2)] The optimal compression ratios $\{R_n^*\}$ are
\begin{equation}
R_n^*=\max\left\{\min\left\{\widehat{R}_n, R_{\rm{max}}\right\}, 1\right\}, \qquad \forall \ n,
\end{equation} where $\widehat{R}_n$ satisfies: $z(\widehat{R}_n)=0$ with the function $z(R_n)$ defined by
\begin{align}\label{Eq:LosslessR}
z(R_n)&=\left[q_n^{(c)}-\frac{1}{g_n f_n}g\left(\frac{1}{d(R_n)R_n}\right)\right]\epsilon e^{\epsilon R_n} \nn\\
&\qquad\qquad-\frac{1}{g_n R_n^2}f'\left(\frac{1}{d(R_n)R_n}\right),
\end{align}
and $d(R_n)\overset{\triangle}{=}\dfrac{T}{\ell_n^{(s)}}-\dfrac{1}{s_n}-\dfrac{1}{f_n}C(R_n,\epsilon)>0$.
\end{itemize}}
\end{proposition}

The proof is given in Appendix~\ref{App:LosslessDataComp}. Though $\widehat{R}_n$ has no closed form, it can be computed by the bisection-search method  in Algorithm~\ref{Al:OptLossless}. This is due to the monotonicity of function $z(R_n)$ as shown in the lemma below and proved in Appendix~\ref{App:Losslessmonotone}.

\begin{algorithm}[t]
  \caption{Optimal Lossless Compression Ratio Algorithm}
  \label{Al:OptLossless}
  \begin{itemize}
\item{\textbf{Step 1} [Initialize]: Let $R_{\ell}=1$, $R_h=R_{\max}$, and $\xi >0$. Obtain $z(R_{\ell})$ and $z(R_h)$.}
\item{\textbf{Step 2} [Bisection search]: \emph{While} $z(R_{\ell}) \neq 0$ and $z(R_h) \neq 0$, update $R_{\ell}$ and $R_h$ as follows.\\
(1) Define $R_n^*=(R_{\ell}+R_h)/2$, compute $z(R_n^*)$.\\
(2) If $z(R_n^*)>0$, let $R_h=R_n^*$, else $R_{\ell}=R_n^*$.\\
\emph{Until} $R_h-R_{\ell}<\xi$.}
\end{itemize}
  \end{algorithm}

\begin{lemma}\label{Lem:Losslessmonotone}
\emph{$z(R_n)$ is a \emph{monotone-increasing} function of $R_n$.}
\end{lemma}

Consequently, the effects of parameters on the optimal compression ratio are detailed below.

\begin{corollary}[Properties of Optimal Compression Ratios]\label{Cor:CompressionRatio}\emph{The optimal compression ratios $\{R_n^*\}$ have the following properties:
\begin{itemize}
\item[\emph{1)}] If the compression powers $\{q_n^{(c)}\}$ are identical but the effective channel power gains satisfy $g_1\ge g_2\cdots \ge g_n$, then $R_1^*\le R_2^* \cdots \le R_N^*$.
\item[\emph{2)}] If the effective channel power gains $\{g_n\}$ are identical but the compression powers satisfy $q_1^{(c)}\ge q_2^{(c)} \cdots  \ge q_n^{(c)}$, then $R_1^*\le R_2^* \cdots \le R_N^*$.
\end{itemize} }
\end{corollary}

Proposition~\ref{Cor:CompressionRatio} is proved in Appendix~\ref{App:CompressionRatio}. Accordingly, it is more energy-efficient for a sensor to perform more aggressive compression in the case of a poor channel or small compression-power consumption. 

\begin{algorithm}[t]
  \caption{Efficient Algorithm for Lossless Compression Solving Problem P1}
  \label{Al:Lossless}
  \begin{itemize}
\item{\textbf{Step 1}: Initialize the compression ratio $R_n^*$.}
\item{\textbf{Step 2}: \emph{Repeat} \\
(1) Given fixed compression ratio, find the optimal power allocation $P_n^*$ and the sensing-data size $(\ell_n^{(s)})^*$ by using Algorithm~\ref{Al:Transmission}.\\
(2) Given fixed sensing-data size, update the compression ratio $R_n^*$ by using Algorithm~\ref{Al:OptLossless}.\\
\emph{Until} convergence or a maximum number of iterations has been reached.}
  \end{itemize}
  \end{algorithm}

\subsection{Complexity Analysis}
Combining the results of the preceding subsections, the efficient control policy for achieving the maximum operator's reward can be computed by an iterative procedure as summarized in Algorithm~\ref{Al:Lossless} with the complexity analyzed as follows.  In the outer iteration, the algorithm alternatively optimizes the control policy given by fixed compression ratios and sensing-data sizes. First, given fixed compression ratios, the optimal control policy can be computed by the one-dimensional search for $\lambda^*$ described in Algorithm~\ref{Al:Transmission}, which invokes an inner iteration. The complexity of the one-dimensional search can be represented by  $\mathcal{O}(\log(1/\xi))$ given the solution accuracy of $\xi>0$. 

Moreover, the complexity of the sensor-transmission, power-allocation, and sensing-data size is $\mathcal{O}(N)$ in each inner iteration. Hence, the total complexity of the control policy given fixed compression ratios is $\mathcal{O}(N\log(1/\xi))$. Next, given fixed sensing-data sizes, the optimal compression ratio for each sensor can be computed by another one-dimensional search in  Algorithm~\ref{Al:OptLossless}. Therefore, each iteration also has the computation complexity of $\mathcal{O}(N\log(1/\xi))$. Last, the complexity of the outer iteration can be characterized by $\mathcal{O}(\log(1/\xi))$. In summary, the total complexity of Algorithm~\ref{Al:Lossless} is $\mathcal{O}(N \log^2(1/\xi))$, which is feasible in practical systems.


\section{Extension to Lossy Compression}
The preceding section considered lossless compression. In this section, those results are extended to the case of lossy compression. To this aim, problem P1 is modified to include the quality factor $b_n=\frac{1}{\sqrt{\check{R}_n}}$ for lossy compression, which leads to its lossy-compression counterpart as follows: 
\begin{equation*}
\begin{aligned}
\max_{\substack{\check{P}_n\ge 0, \check{\ell}_n^{(s)} \ge 0,\\ \check{R}_n\in [1, \check{R}_{\rm{max}}], \check{t}_n\ge 0 }} \quad 
&\sum_{n=1}^{N} a_n \log\left(1\!+\!\frac{1}{\sqrt{\check{R}_n}}\check{\ell}_n^{(s)}\right)\!-\!c\sum_{n=1}^{N} \check{P}_n T_0\\ 
\text{s.t.} \qquad 
&\sum_{n=1}^{N} \check{P}_n \le P_0,\\
\textbf{(P3)}\qquad \qquad  
&\frac{\check{\ell}_n^{(s)}}{s_n}+\frac{\check{\ell}_n^{(s)} (e^{\check{\epsilon} \check{R}_n}\!-\!e^{\check{\epsilon}})}{f_n}+\check{t}_n\le T,\quad~~~ \forall~n, \\
&\check{\ell}_n^{(s)} \left[q_n^{(r)}+q_n^{(s)}+q_n^{(c)}(e^{\check{\epsilon}\check{R}_n}-e^{\check{\epsilon}})\right] \\
&\qquad\quad+\dfrac{\check{t}_n}{g_n}\left(\frac{\check{\ell}_n^{(s)}}{\check{R}_n \check{t}_n}\right) \le \eta g_n \check{P_n} T_0, ~~ \forall~n.
\end{aligned}
\end{equation*}

Problem P3 has a similar form as P1 and it is also non-convex. Leveraging the  iterative solution approach developed in the preceding section, problem P3 can be decomposed into two sub-problems for optimizing two corresponding subsets of parameters, $\{\check{P}_n, \check{\ell}_n^{(s)}, \check{t}_n\}$ and $\{\check{R}_n, \check{\ell}_n^{(s)}, \check{t}_n\}$. The first sub-problem has the same form as problem P1-A in the lossless-compression counterpart; thus, it can be solved in a similar way (see Section~\ref{Sec:JointFixedCompress}). The second sub-problem (see problem P4 below) differs from problem P1-B due to the addition of sensing-data sizes $\{\check{\ell}_n^{(s)}\}$ to the optimization variables. The reason is that lossy-compression causes information loss and thereby reduces data utility, which has to be compensated for by increasing the sensing-data sizes. To address this issue and also for better tractability, we apply the constraint that the data utility contributed by different sensors is fixed in problem P4 as $\{u_n\}$, which results from solving the first sub-problem. Consequently, problem P4 follows from the problem P3 as: 
\begin{equation}
\text{(Data utility constraint)}~~u_n=b_n\check{\ell}_n^{(s)}=\frac{1}{\sqrt{\check{R}_n}}\check{\ell}_n^{(s)}.
\end{equation}
Then, the corresponding optimization problem can be formulated as follows.
\begin{equation*}
\begin{aligned}
\min_{\substack{\check{R}_n\in [1,\check{R}_{\rm{max}}],\\ \check{t}_n\ge 0, \check{\ell}_n^{(s)}\ge0}}~
&\check{\ell}_n^{(s)} \left[q_n^{(r)}\!+\!q_n^{(s)}\!+\!q_n^{(c)} \left(e^{\check{\epsilon} \check{R}_n}\!-\!e^{\check{\epsilon}}\right)\right]\!+\!\frac{\check{t}_n}{g_n}f\left(\frac{\check{\ell}_n^{(s)}}{\check{R}_n \check{t}_n}\right)\\
\text{s.t.} \qquad 
&\frac{1}{\sqrt{\check{R}_n}}\check{\ell}_n^{(s)}=u_n,\qquad\qquad\qquad\qquad\qquad~~~ \forall~n,\\
\textbf{(P4)} \qquad
&\check{t}_n+\frac{\check{\ell}_n^{(s)}}{s_n}+\frac{\check{\ell}_n^{(s)}}{f_n}\left(e^{\check{\epsilon} \check{R}_n}-e^{\check{\epsilon}}\right)\leq T, \qquad~~~~ \forall~n.
\end{aligned}
\end{equation*}
This problem is solved in the following sub-section. 

\subsection{Joint Optimization of Sensing-Data Sizes, Compression, and Transmission}
Problem P4 can be shown to be non-convex but may also be transformed into a convex formulation as follows. Here, we define a set of new variables $\{r_n\}$ as $r_n=\sqrt{\check{R}_n}\in \left[1, r_{\rm{max}}\right]$, where $r_{\rm{max}}=\sqrt{\check{R}_{\rm{max}}}$. Hence, $\check{\ell}_n^{s}=u_n r_n$ and $\check{R}_n=r_n^2$. Substituting them into problem P4 yields:
\begin{equation*}
\begin{aligned}
\min_{\substack{r_n\in\left[1, r_{\rm{max}}\right],\\ \check{t}_n\ge 0}}~
&u_n r_n \left[q_n^{(r)}\!+\!q_n^{(s)}\!+\!q_n^{(c)}\left(e^{\check{\epsilon}r_n^2}\!-\!e^{\check{\epsilon}}\right)\right]\!+\!\frac{\check{t}_n}{g_n}f\left(\frac{u_n}{\check{t}_n r_n}\right)\\
\textbf{(P5)}~~\text{s.t.}~~
&\check{t}_n+u_n r_n\left(\frac{1}{s_n}+\frac{e^{\check{\epsilon}r_n^2}-e^{\check{\epsilon}}}{f_n}\right) \leq T, \qquad\qquad \forall~n.
\end{aligned}
\end{equation*}
Using a similar approach as that for deriving Lemma~\ref{Lem:losslessconv}, it can be proved that problem P5 is a convex optimization problem. Hence, applying the Lagrange method as for solving problem P1-B, the key results of this subsection are derived as stated in the following corollary.

\begin{proposition}[Optimal Lossy Compression and Transmission]\label{Pro:LossyDataComp}\emph{The solution of P5 is as follows. 
\begin{itemize}
\item[1)] The optimal sensor-transmission durations $\{\check{t}_n^*\}$ are
\begin{equation}
\check{t}_n^*= T - \dfrac{u_n r_n^*\left(e^{\check{\epsilon}{r_n^*}^2}-e^{\check{\epsilon}}\right)}{f_n} -\dfrac{u_n r_n^*}{s_n}, \qquad \forall \ n,
\end{equation}
which corresponds to full utilization of the crowd-sensing duration.
\item[2)] The optimal square roots of compression ratios $\{r_n^*\}$ are
\begin{equation}
r_n^*=\max\left\{\min\left\{\widehat{r}_n, r_{\rm{max}}\right\}, 1\right\}, \qquad \forall \ n,
\end{equation} where $\widehat{r}_n$ satisfies: $\check{z}(\widehat{r}_n)=0$ with the function $\check{z}(r_n)$ defined by
\begin{align}
\!\!\check{z}(r_n)&\!=\!r_n Q'(r_n)\!+\!Q(r_n)\!-\!\frac{1}{g_n r_n^2}f'(\frac{1}{\check{d}(r_n)r_n})\nonumber\\
&\qquad\!-\!\frac{1}{g_n}g\left(\frac{1}{\check{d}(r_n)r_n}\right)\left[V(r_n)\!+\!r_n V'(r_n)\right],
\end{align}
where $Q(r_n)\!=\!q_n^{(r)}\!+\!q_n^{(s)}\!+\!q_n^{(c)}\left(e^{\check{\epsilon}r_n^2}\!-\!e^{\check{\epsilon}}\right)$, $V(r_n)\!=\!\left(\frac{1}{s_n}\!+\!\frac{e^{\check{\epsilon}r_n^2}\!-\!e^{\check{\epsilon}}}{f_n}\right)$ and $\check{d}(r_n)=\frac{T}{u_n}\!-\!r_n\left(\dfrac{1}{s_n}\!+\!\frac{e^{\check{\epsilon}r_n^2}\!-\!e^{\check{\epsilon}}}{f_n}\right)$.
\end{itemize}}
\end{proposition}


This proposition can be proved by a similar method as that for deriving Proposition~\ref{Pro:LosslessDataComp}. 
Even though $\widehat{r}_n$ has no closed form, it can be computed by a bisection method due to the monotonicity  of function $\check{z}(r_n)$ as shown in the lemma below, which is proved in Appendix~\ref{App:Lossymonotone}.

\begin{lemma}\label{Lem:Lossymonotone}
\emph{$\check{z}(r_n)$ is a \emph{monotone-increasing} function of $r_n$.}
\end{lemma}

Following a similar approach as that for deriving Corollary~\ref{Cor:CompressionRatio}, it can be proved that the optimal compression ratio increases with the effective channel power gain and decreases with the compression power.

\begin{remark}[Lossy vs. Lossless Compression]\emph{Given the same sensing-data size, as compared to lossless compression, the lossy operation consumes less energy for compression and results in smaller size of compressed data, which further reduces transmission-energy consumption. The disadvantage, however, is that it sacrifices a part of data utility due to information loss. Hence, there exists a tradeoff between the energy consumption and the data utility for lossy compression. The decision of choosing lossless or lossy compression is intractable due to the lack of closed-form expression for the maximum reward (see Proposition~\ref{Pro:LosslessDataComp} and Corollary~\ref{Pro:LossyDataComp}). However, it can be observed that lossy compression is preferred under the following conditions: poor channel, high compression power, or low sensing power. In these cases, lossy compression can achieve significant energy savings.
}
\end{remark}



\section{Simulations and Discussion}

The simulation parameters are set as follows unless specified otherwise. Our WPCS system comprises $1$ AP equipped with $N_t=40$ antennas and $10$ MSs each with a single antenna. The $N_t$ antennas are divided into $10$ groups, each having $N_t/10=4$ antennas and pointing to one specific MS by transmit beamforming. The time duration of the WPT phase is $T_0=1$ s. For each MS $n$, the channel vector $\bold{h_n} \in \mathbb{C}^{4\times 1}$ is assumed to follow Rician fading and thus is modeled as $$\bold{h_n}=\sqrt{\frac{\Omega_n K}{1+K}}\bold{\bar{h}_n}+\sqrt{\frac{\Omega_n}{1+K}}\bold{\widehat{h}_n}, $$ where the Rician factor $K=10$, the average fading power gain $\Omega_n= 5\times 10^{-4}\times {d_n}^{-2}$, the distance between the AP and MS $d_n$ is distributed uniformly in the range $[1,5]$ m, the \emph{line-of-sight} (LOS) component $\bold{\bar{h}_n}$ has all elements equal to one, and $\bold{\widehat{h}_n}$ is a $4\times 1$ i.i.d. $\mathcal{CN}(0,1)$ vector representing small-scale fading. The effective channel power gain $g_n=|\bold{h_n}|^2$ as follows from transmit/receive beamforming. The energy conversion efficiency is set as $\eta=0.5$. 

Further, the effective bandwidth $B=10$ KHz and the variance of complex-white-Gaussian-channel noise $N_0=10^{-9}$ W. For the system reward, the utility weights for all the MSs are the same: $a_n=0.04$ and the cost weight $c=0.6$. For each MS, the sensing rate follows a uniform distribution with $s_n \in [1,10] \times 10^4$ bits/s and the energy consumption per bit is distributed uniformly by $q_n^{(s)} \in [1,10] \times 10^{-12}$ J/bit \cite{thiagarajan2011accurate}. For the data compression, the required number of CPU cycles per bit, CPU-cycle frequency, and energy consumption per cycle follow the uniform distribution with $C_n \in [0,3000]$ cycles/bit, $f_n \in [0.1,1]$ GHz, and $q_n^{(c)} \in [1,10] \times 10^{-14}$ J/cycle, respectively. In addition, the energy reward per bit is distributed uniformly as: $q_n^{(r)} \in [1,10] \times 10^{-12}$ J/bit. The maximum compression ratios for lossless and lossy compression are set as $R_{\max} = 3$ and $\check{R}_{\max} = 25$ with $\epsilon = 4$ and $\check{\epsilon} = 0.1$, respectively.

\subsection{Wireless Power Allocation}

\begin{figure}[t]
  \centering
  \includegraphics[scale=0.38]{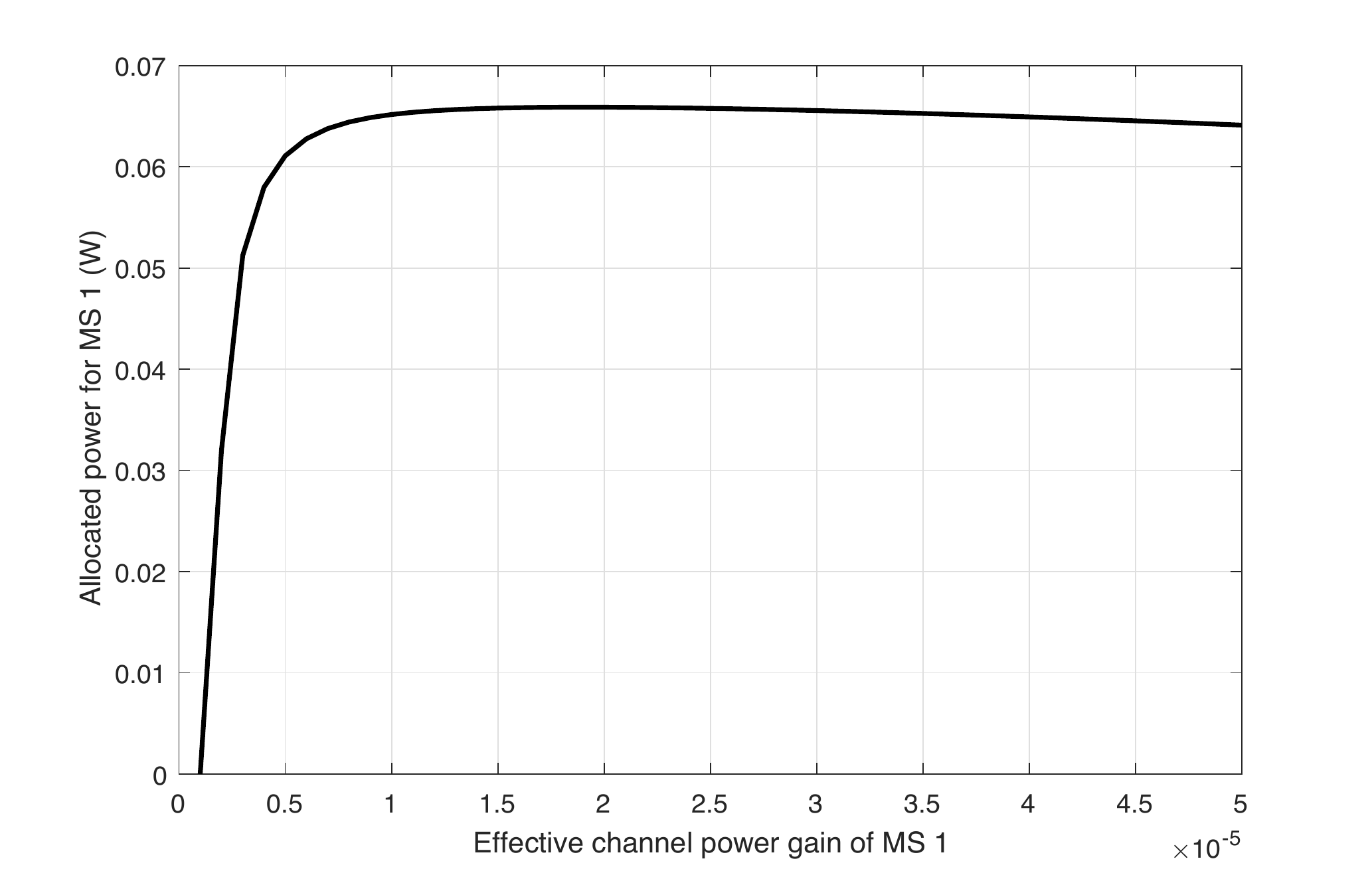}
  \caption{Impact of effective power channel gain on allocated wireless power.}
  \label{Fig:PowerVSChannel}
\end{figure}

Numerical results produced by using Proposition~\ref{Pro:LosslessPowerAllo} are presented in Fig.~\ref{Fig:PowerVSChannel} to offer insights into the optimal policy for wireless-power allocation. The effective channel power gain of sensor $1$ varies, while those of sensors $2,\cdots,N$ are fixed as $10^{-5}$. One can observe that if the channel is poor, sensor $1$ does not receive wireless power due to the small priority function. Otherwise, if the effective channel power gain exceeds a threshold (of about $10^{-6}$), the allocated wireless power first increases and then decreases. This curious observation indicates that \emph{when the effective channel power gain is not sufficiently large, the sensor should receive enough power for increasing its data utility}. However, equipped with a good channel, the sensor can contribute to large data utility even with smaller received power, which allows for saving more power for other sensors in order to increase the sum data utility.

\subsection{Iteration Convergence Rate}
\begin{figure}[t]
  \centering
  \subfigure[Operator's reward vs. iteration]{
  \label{FigIteration}
  \includegraphics[scale=0.38]{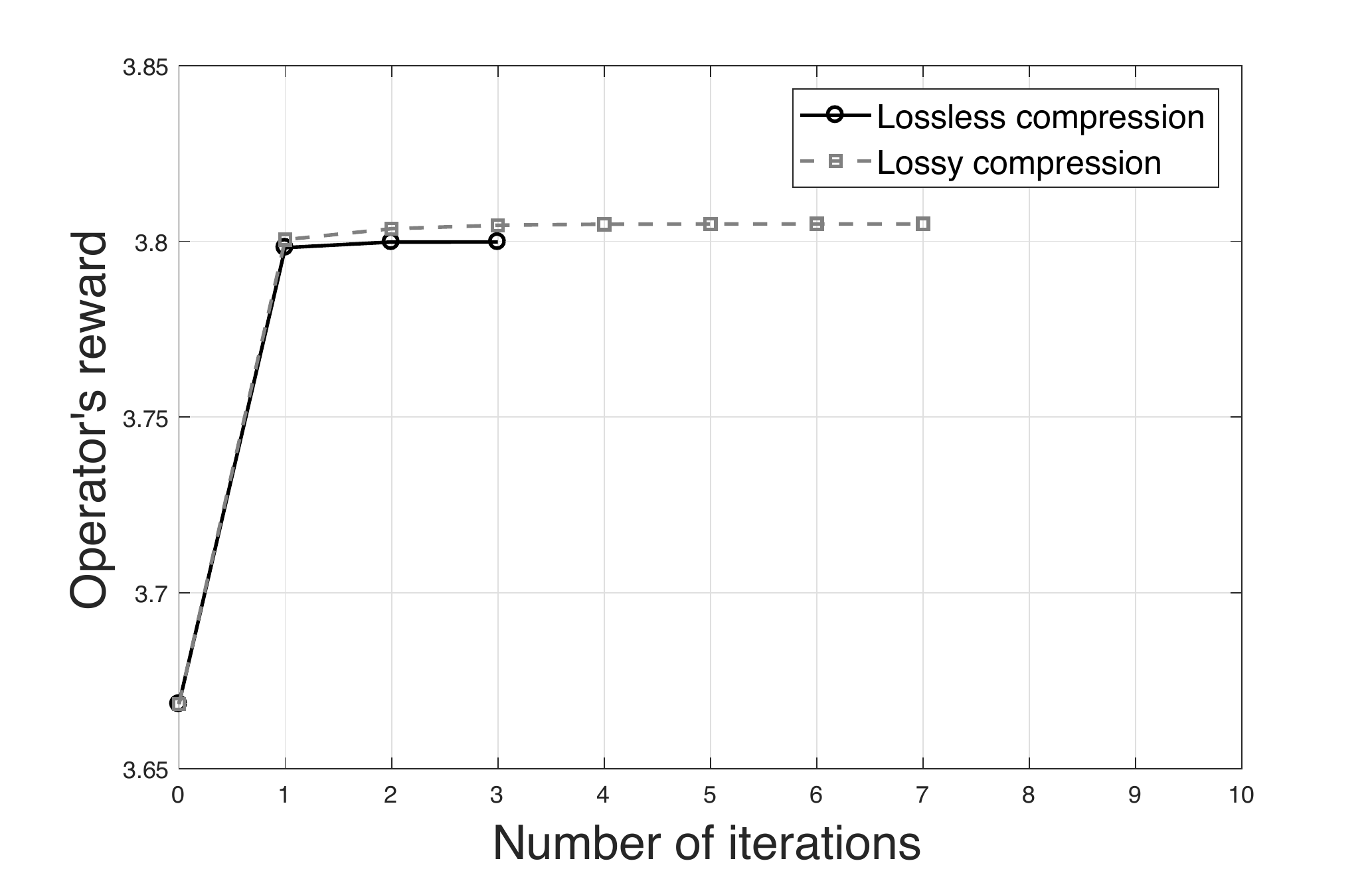}}
  \subfigure[Reward gap vs. iteration]{
  \label{FigGap}
  \includegraphics[scale=0.38]{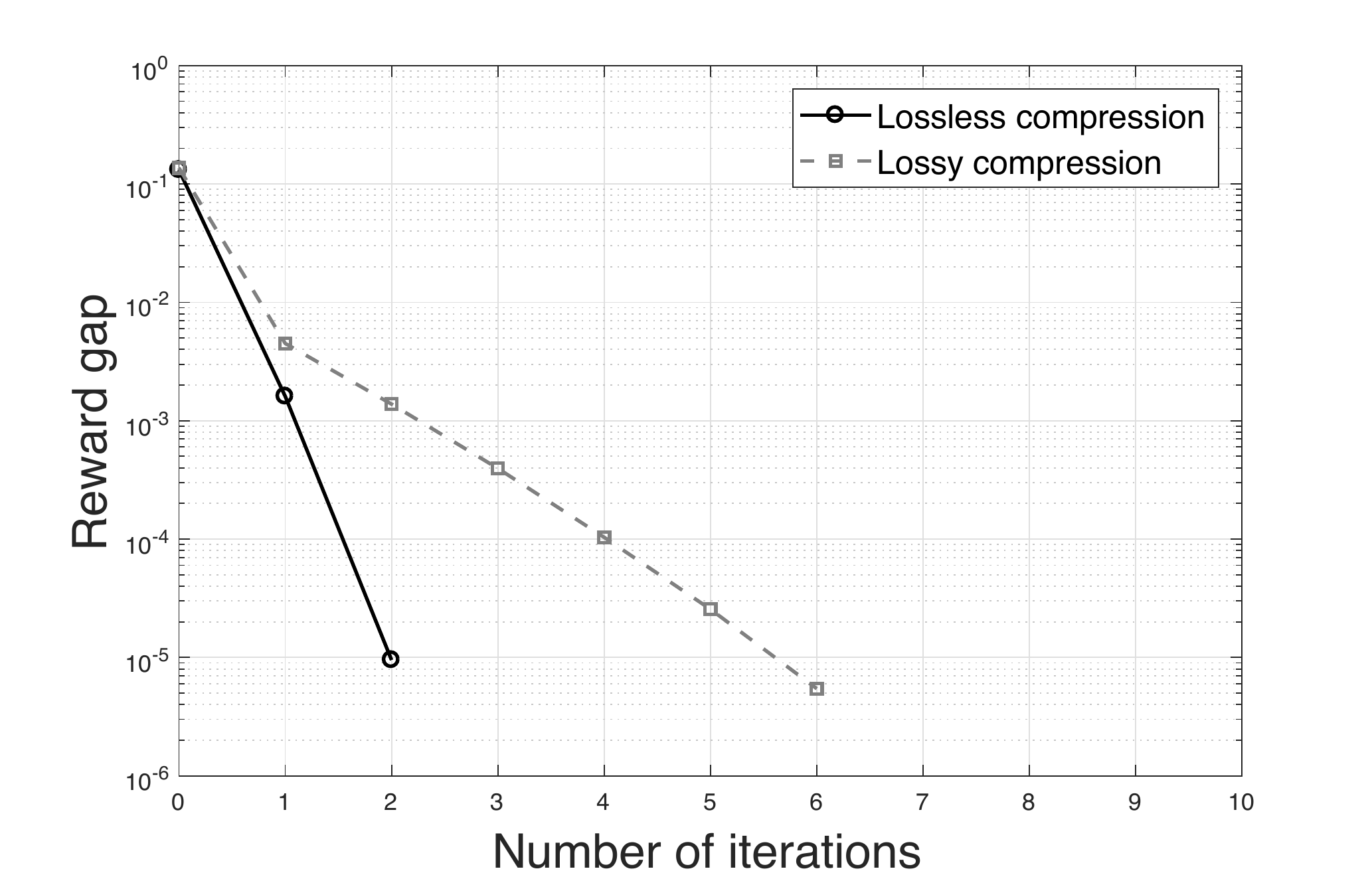}}
  \caption{Convergence of our proposed algorithm.}
  \label{FigConverge}
\end{figure}

The convergence of our proposed algorithm is assessed in Fig.~\ref{FigConverge}, where the allowed error is set as $\xi=10^{-5}$. Observe from Fig.~\ref{FigIteration} that the operator's reward increases monotonically as the iteration evolves and tends to converge after only a few iterations (about $3$ times for lossless compression and $7$ for lossy compression). Next, let $R(k)$ denote the operator's reward after $k$-th iteration and $R^*$ is the maximum operator's reward of our proposed algorithm. Define $R_{\rm{gap}}$ as the reward gap between $R(k)$ and $R^*$ given as  $R_{\rm{gap}}=|R(k)-R^*|$. Fig.~\ref{FigGap} displays the curves of the reward gap verses the number of iterations. We can observe that the gap of our proposed algorithm almost \emph{linearly} decreases as the iteration evolves, thus our proposed solution can be approximated as a first-order algorithm with linear local convergence rate \cite{schatzman2002numerical}. 

\subsection{WPCS Performance}
Three baseline policies are considered for our performance comparison. The first one is the \emph{baseline policy with fixed compression ratio} (FCR) that jointly optimizes the power allocation, sensor transmission, and sensing-data size by following the approach in Section~\ref{Sec:JointFixedCompress}. Specifically, the compression ratios for lossless and lossy compression are set as $R_n=1.5$ and $\check{R}_n=4$, respectively, which can achieve the maximum operator's reward for the baseline policy with FCR. The second one is the \emph{baseline policy with equal power allocation} (EPA), which jointly optimizes the sensor transmission, sensing-data size, and compression ratio given the power allocation policy. The third one is the \emph{baseline policy without compression}, which can be regarded as a special case of the FCR with the compression ratio set as zero.

\begin{figure}[t]
  \centering
  \subfigure[Lossless compression]{
  \label{FiglosslessE}
  \includegraphics[scale=0.38]{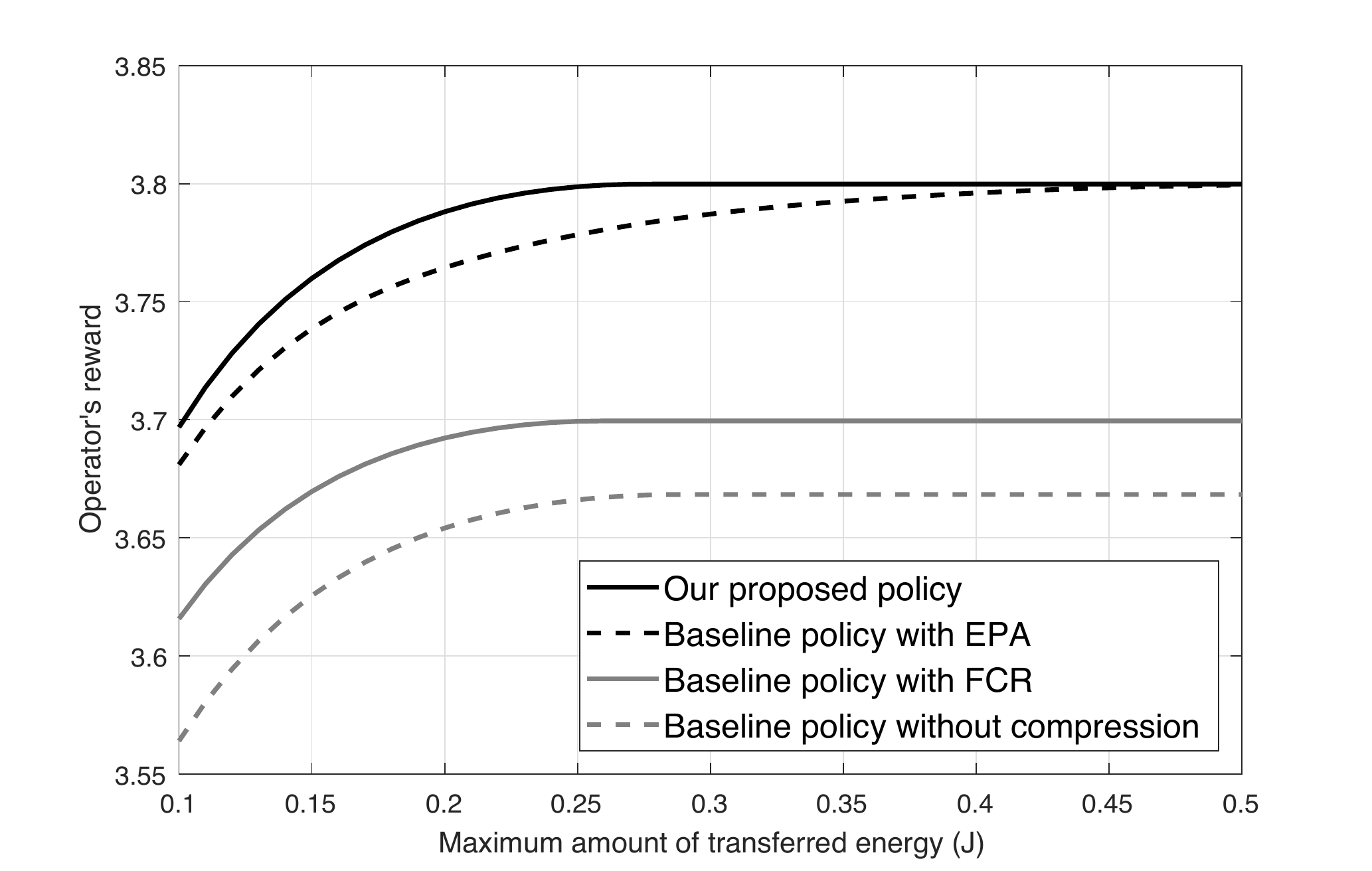}}
  \subfigure[Lossy compression]{
  \label{FiglossyE}
  \includegraphics[scale=0.38]{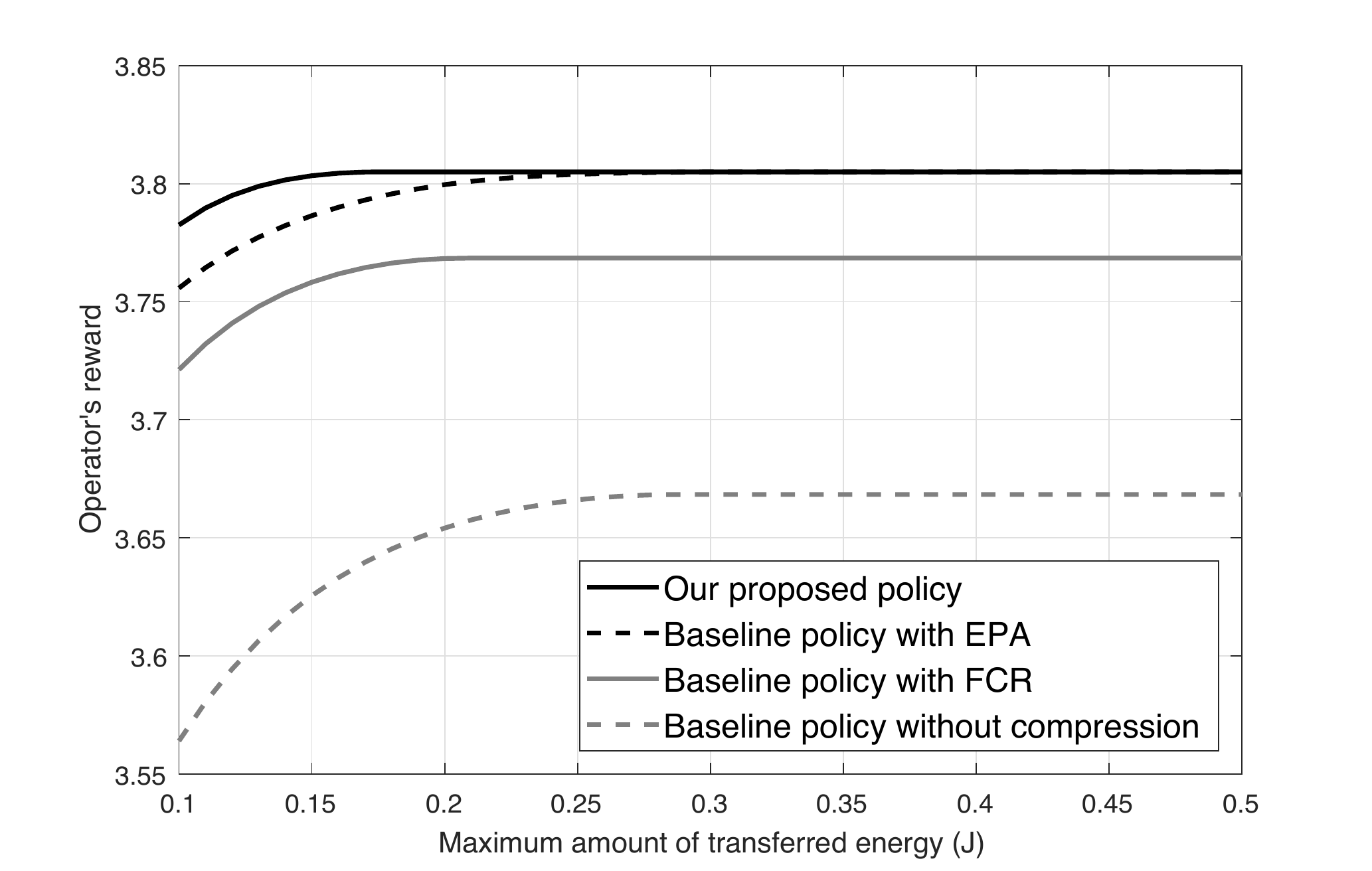}}
  \caption{Operator's reward vs. maximum amount of transferred energy.}
  \label{FigRewardE}
\end{figure}

\begin{figure}[t]
  \centering
  \subfigure[Lossless compression]{
  \label{FiglosslessT}
  \includegraphics[scale=0.38]{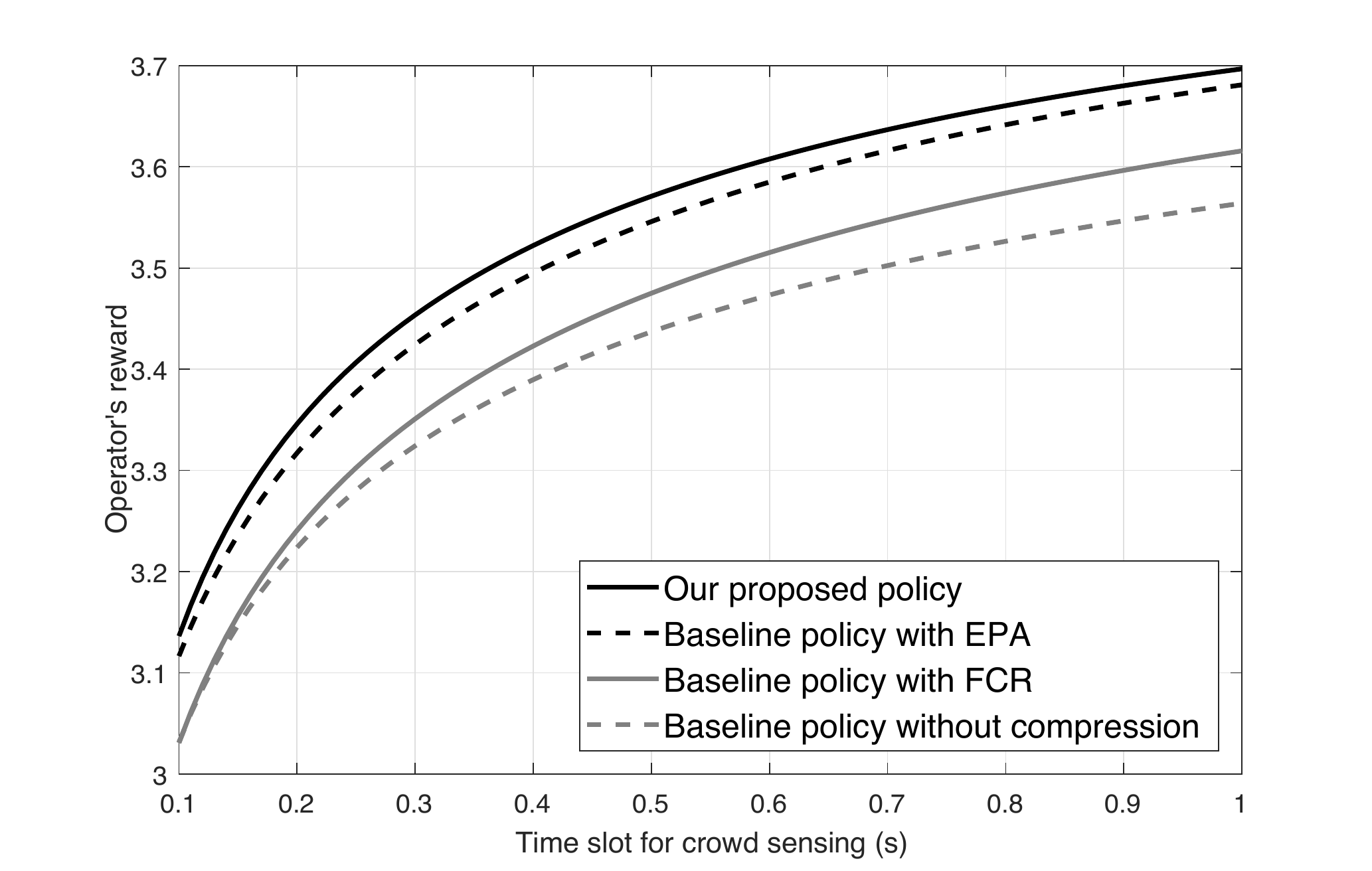}}
  \subfigure[Lossy compression]{
  \label{FiglossyT}
  \includegraphics[scale=0.38]{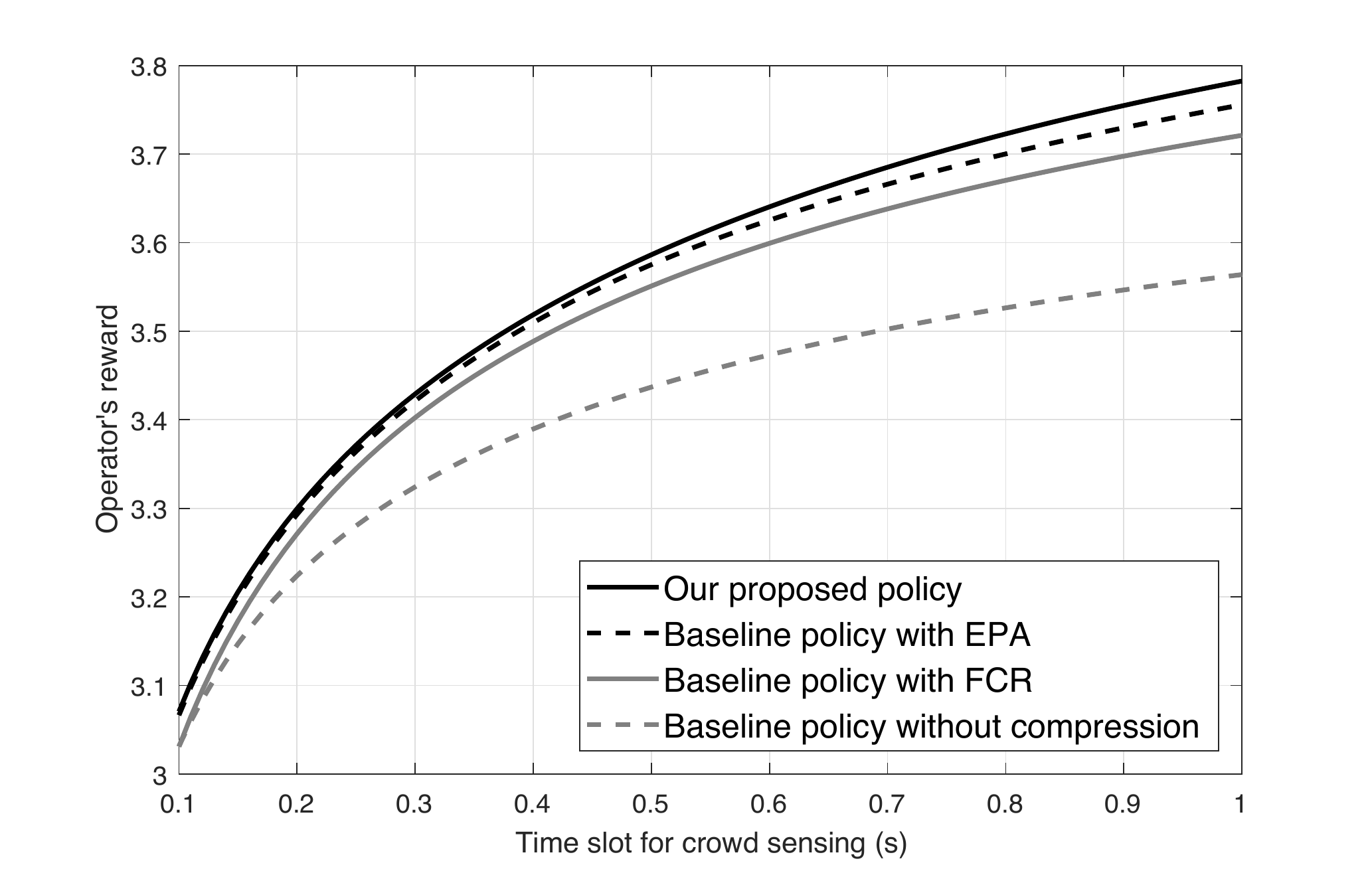}}
  \caption{Operator's reward vs. time slot for crowd sensing.}
  \label{FigRewardT}
\end{figure}

The curves for the operator's reward versus the maximum amount of transferred energy are displayed in Fig.~\ref{FigRewardE} with the time duration for crowd sensing set as $T=1$ s. For lossless compression, it can be observed from Fig.~\ref{FiglosslessE} that the operator's reward for our proposed policy is monotone-increasing with the growth of transferred energy in the small transferred-energy regime. However, when the amount of energy exceeds a threshold (of about $0.25$ J), the performance cannot improve further by increasing the transferred energy. The reason is that the bottleneck in this case is no longer energy but other settings, such as the crowd-sensing duration. 

Note that \emph{our proposed policy has significant performance gain over the policy with FCR and without compression}, even when the maximum amount of transferred energy is sufficient. This shows the necessity of optimizing the compression ratios to improve system performance when the transferred energy is sufficient. In addition, among the three baseline policies, \emph{the one with EPA has the highest operator's reward and approaches close-to-optimal performance in the large transferred energy regime}. This is due to the fact that optimizing compression ratios can substantially reduce transmission energy consumption, thus allowing for using more energy to sense data for better data utility. For lossy compression, one can observe from Fig.~\ref{FiglossyE} that our proposed policy yields higher operator's reward with respect to the lossless counterpart for the case of small and large transferred energy, respectively. Other observations are similar to those in Fig.~\ref{FiglosslessE}.

Fig.~\ref{FigRewardT} demonstrates the curves of the operator's reward versus the crowd-sensing duration for both compression methods. One can observe from Fig.~\ref{FiglosslessT} that for lossless compression, the operator's reward is monotone-increasing with the extension of crowd-sensing duration, as it can reduce transmission-energy consumption so as to save more energy for data sensing. However, the growth rate of the operator's reward slows down with the crowd-sensing duration. It indicates that \emph{extending a short crowd-sensing duration can substantially increase the operator's reward.} Moreover, in long crowd-sensing duration regime, lossy compression yields higher operator's reward than in the lossless case. Other observations are similar to those in Fig.~\ref{FigRewardE}.

\begin{figure}[t]
  \centering
  \includegraphics[scale=0.38]{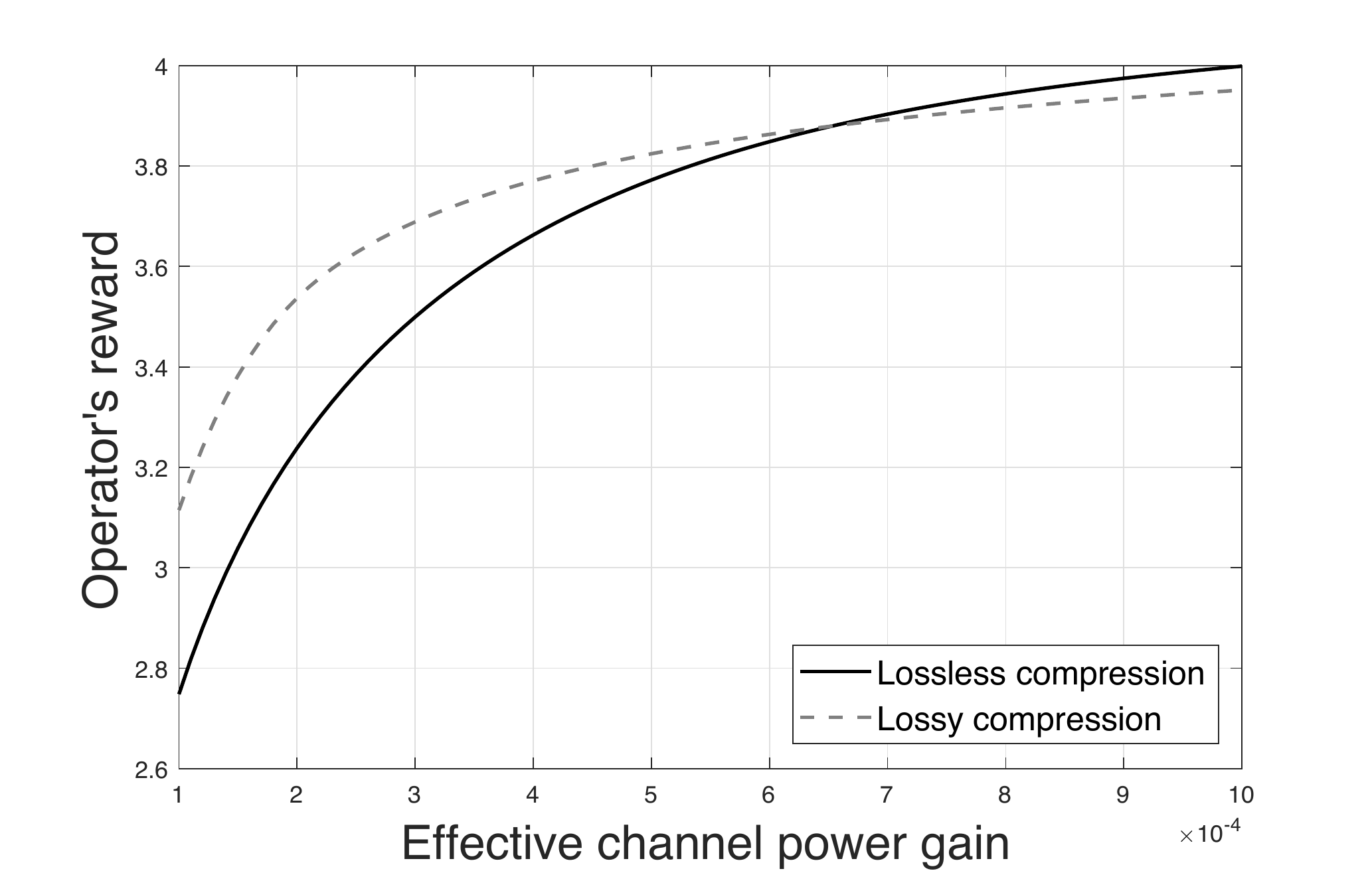}
  \caption{Operator's reward vs. effective channel power gain.}
  \label{FigRewardh}
\end{figure}

Finally, the impact of effective channel power gain on the compression method selection is evaluated in Fig.~\ref{FigRewardh}. Several observations are made as follows. First, the operator's reward is observed to be a monotone-increasing function of the effective channel power gain. Next, \emph{lossy and lossless compression is preferred for relatively small and large effective channel power gains, respectively}. The reason is that for poor channels, transmission-energy can dominate other components (e.g., sensing), and it can be substantially reduced by lossy compression due to efficient data-compression performance. However, when the channel is good enough, lossless compression can contribute to higher data utility with small compression energy, and thereby may be preferred over the lossy operation.

\section{Concluding Remarks}
This work proposed a multiuser WPCS system model and investigated a joint control of power allocation, sensing, compression, and transmission for maximizing the operator's reward. It offered several insights into the control policies of the operator and the MSs. To be specific, for the operator, we showed that it should select the MSs for participating in the crowd sensing operation based on its configuration and the individual parameters of MSs. In addition, the optimal power allocation policy has a threshold-based structure depending on the derived crowd sensing priority function. For MSs, each selected sensor should control its compression ratio, sensor transmission, and sensing-data size for minimizing the resultant energy consumption. This work can be further extended to other interesting scenarios, such as cooperation among multiple operators and online control of multiuser WPCS systems.
%
%
 
\appendix
\subsection{Proof of Lemma~\ref{Lem:NecCond}}\label{App:NecCond}
This lemma can be proved by contradiction. First, assume that $\left\{\left(\ell_n^{(s)}\right)^*, t_n^*, P_n^*\right\}$ is the optimal solution satisfying $\alpha_n \left(\ell_n^{(s)}\right)^*\!+\!\frac{t_n^*}{g_n} f\left(\frac{\left(\ell_n^{(s)}\right)^*}{R_n t_n^*}\right) \!<\! \eta g_n P_n^* T_0$. Then, there exists another feasible power allocation policy, denoted by $P_n^{(1)}$, which satisfies $P_n^{(1)}<P_n^{*}$ and $\alpha_n {(\ell_n^{(s)})}^*\!+\!\frac{t_n^*}{g_n} f\left(\frac{\left(\ell_n^{(s)}\right)^*}{R_n t_n^*}\right) \!=\! \eta g_n P_n^{(1)} T_0$. It leads to the following 
\begin{equation*}
\begin{aligned}
&\sum_{n=1}^N a_n \log(1+(\ell_n^{(s)})^* )-c\sum_{n=1}^N P_n^{(1)} T_0 \\
&\qquad\qquad > \sum_{n=1}^N a_n \log(1+(\ell_n^{(s)})^*)-c\sum_{n=1}^N P_n^* T_0,
\end{aligned}
\end{equation*}
which contradicts the optimality. Similarly, if the optimal solution leads to $\beta_n \left(\ell_n^{(s)}\right)^*+t_n^* < T$, then there exists another feasible policy with $(\ell_n^{(s)})^{(1)}$ that satisfies $(\ell_n^{(s)})^{(1)}=\frac{T-t_n^*}{\beta_n}>\left(\ell_n^{(s)}\right)^*$. This new policy yields higher operator's reward as
\begin{equation*}
\begin{aligned}
&\sum_{n=1}^N a_n \log(1+(\ell_n^{(s)})^{(1)})-c\sum_{n=1}^N P_n^* T_0 \\
&\qquad\qquad > \sum_{n=1}^N a_n \log(1+\left(\ell_n^{(s)}\right)^*)-c\sum_{n=1}^N P_n^* T_0,
\end{aligned}
\end{equation*}
thus contradicting the assumption. Combining the above together leads to the desired result.

\subsection{Proof of Lemma~\ref{Lem:conv}}\label{App:conv}
First, it is easy to see that $\sum_{n=1}^N a_n \log\left(1+\frac{T-t_n}{\beta_n}\right)$ is concave and $\frac{\alpha_n(T-t_n)}{\eta g_n \beta_n}$ is linear for $t_n$. Then, the second derivative of $\frac{t_n}{\eta g_n^2} f\left(\frac{T-t_n}{{R_n}{\beta_n}{t_n}}\right)$ is $\frac{N_0{(T\ln2)}^2}{\eta (g_n B R_n \beta_n)^2 t_n^3}2^\frac{T-t_n}{B R_n \beta_n t_n}>0$ for $0<{t_n}<T$; thus, it is a convex function. It follows that the objective function, being a summation over a set of convex functions, preserves the convexity. Combining it with the convex constraints yields the desired result.

\subsection{Proof of Corollary~\ref{Cor:StruTime}}\label{App:StruTime}
From \eqref{Eq:LosslessOpt}, the utility weight can be expressed as:
\begin{equation*}
\begin{aligned}
a_n &= \left[\left(\frac{T\ln2}{B R_n \beta_n t_n^*}-1\right)e^{\frac{T\ln2}{B R_n \beta_n t_n^*}-\frac{\ln2}{B R_n \beta_n}}+\frac{\alpha_n g_n}{N_0 \beta_n}+1\right]\\
&\qquad\qquad \times \frac{(\lambda^*+c)N_0[\beta_n+(T-t_n^*)]}{\eta g_n^2}.
\end{aligned}
\end{equation*}
Given $a_1<a_2$, it can be directly observed from the above equation that $t_1>t_2$. As for the effective channel power gain, it should satisfy the following equation:
\begin{align}\label{Eq:ProofCor1}
&g_n \left[\frac{a_n \eta g_n}{(\lambda^*+c)N_0[\beta_n+(T-t_n^*)]}-\frac{\alpha_n}{N_0 \beta_n}\right]\nonumber\\
& \qquad\qquad = \left(\frac{T\ln2}{B R_n \beta_n t_n^*}-1\right)e^{\frac{T\ln2}{B R_n \beta_n t_n^*}-\frac{\ln2}{B R_n \beta_n}}+1.
\end{align}
Note that both the right hand-side of \eqref{Eq:ProofCor1} and $N_0[\beta_n+(T-t_n^*)]$ are decreasing with $t_n$. It means that if $g_1<g_2$, it should be satisfied that $t_1\ge t_2$, thus completing the proof.


\subsection{Proof of Proposition~\ref{Pro:LosslessPowerAllo}}\label{App:LosslessPowerAllo}
First, combing $\alpha_n \left(\ell_n^{(s)}\right)^*+\frac{t_n^*}{g_n} f\left(\frac{\left(\ell_n^{(s)}\right)^*}{R_n t_n^*}\right) = \eta g_n P_n^* T_0$ and $\beta_n \left(\ell_n^{(s)}\right)^*+t_n^*= T$ gives
\begin{equation*}
P_n^*=\frac{1}{\eta g_n T_0}\left[\frac{t_n^*}{g_n}f\left(\frac{T-t_n^*}{R_n \beta_n t_n^*}\right)+\frac{\alpha_n (T-t_n^*)}{\beta_n}\right].
\end{equation*}

Next, the first and the second derivative of $P_n$ are calculated as follows.
\begin{align*}
&\frac{\partial P_n}{\partial t_n}\!=\!\frac{N_0}{\eta g_n^2 T_0}\left[\left(1-\frac{T\ln2}{B R_n \beta_n t_n}\right)2^\frac{T-t_n}{B R_n\beta_n t_n}\!-\!1\right]\!-\!\frac{\alpha_n}{\eta \beta_n g_n T_0}, \\
&\frac{\partial^2 P_n}{\partial t_n^2}\!=\!\frac{N_0{(T\ln2)}^2}{\eta T_0 (g_n B R_n \beta_n)^2 t_n^3}2^\frac{T-t_n}{B R_n \beta_n t_n}.
\end{align*}
It can be observed that $\frac{\partial^2 P_n}{\partial t_n^2}>0$ for $0\le t_n\le T$. Hence, $\frac{\partial P_n}{\partial t_n}$ is monotone-increasing for $0\le t_n\le T$ and 
$$\left(\frac{\partial P_n}{\partial t_n}\right)_{\rm{max}}\!=\!\frac{\partial P_n}{\partial t_n}\bigg|_{t_n=T}\!=\!-\frac{N_n \ln2}{\eta g_n^2 B R_n \beta_n T_0}\!-\!\frac{\alpha_n}{\eta \beta_n g_n T_0}\!<\!0.$$ 
It means that $P_n$ is monotone decreasing with $t_n$ and  $({P_n})_{\rm{min}}={P_n}|_{t_n=T}=0$.
Combining  \eqref{Eq:KKT1} and $0\le t_n\le T$ yields 
\begin{equation*}
\begin{aligned}
\lambda^*&>\frac{\frac{a_n \eta g_n }{\beta_n+(T-t_n^*)}}{\frac{\alpha_n}{\beta_n}-\frac{N_0}{g_n}\left[\left(1-\frac{T\ln2}{B R_n \beta_n t_n^*}\right)2^\frac{T-t_n^*}{B R_n \beta_n t_n^*}-1\right]}-c \\
&=\frac{a_n \eta g_n}{q_n^{(r)}+q_n^{(s)}+q_n^{(c)} e^{\epsilon R_n}+\frac{N_0\ln2}{g_n B R_n}}-c = \phi_n.\end{aligned}
\end{equation*}
If $\lambda^*<\phi_n$, the optimal sensing time will violate the time constraint $t_n^* \!>\! T$; thus, the sensor  will not be selected and no power will be allocated to it i.e., $P_n=0$, which completes the proof.


\subsection{Proof of Lemma~\ref{Lem:CompOrNot}}\label{App:CompOrNot}
Based on \eqref{Eq:LosslessKKT1}, for $R_n^*=1$, it can be established that
$\frac{N_0 \ln2}{g_n B}2^{\frac{\ell_n^{(s)}}{t_n^* B}}-\left(q_n^{(c)}+\frac{\widehat{\lambda}^*}{f_n}\right){\epsilon}e^{\epsilon}<0.$ Combining it with the optimal conditions, $t_n^*=T-\frac{\ell_n^{(s)}}{s_n}$ and $\widehat{\lambda}^*=\frac{N_0}{g_n}\left[\left(\frac{\ell_n^{(s)} \ln2}{t_n^* B}-1\right)2^\frac{\ell_n^{(s)}}{t_n^* B}+1\right]$, leads to the desired result.

\subsection{Proof of Proposition~\ref{Pro:LosslessDataComp}}\label{App:LosslessDataComp}
Based on \eqref{Eq:LosslessKKT3}, at least one of the two equations $\widehat{\lambda}^*=0$ or $t_n^*+\frac{\ell_n^{(s)}\left(e^{\epsilon R_n^*}-e^{\epsilon}\right)}{f_n}+\frac{\ell_n^{(s)}}{s_n}-T=0$ should be satisfied. Assume that $t_n^*+\frac{\ell_n^{(s)}\left(e^{\epsilon R_n^*}-e^{\epsilon}\right)}{f_n}+\frac{\ell_n^{(s)}}{s_n}-T \neq 0$, then it has $\widehat{\lambda}^*=0$, which leads to  $R_n^* \to \infty$ according to \eqref{Eq:LosslessKKT2}. This, however, contradicts the finite compression ratio. Hence, it should satisfy:  $t_n^*+\frac{\ell_n^{(s)}\left(e^{\epsilon R_n^*}-e^{\epsilon}\right)}{f_n}+\frac{\ell_n^{(s)}}{s_n}=T$. The optimal solution should satisfy \eqref{Eq:LosslessKKT1} as  
\begin{equation*}
\left(q_n^{(c)}+\frac{\widehat{\lambda}^*}{f_n}\right)\epsilon e^{\epsilon R_n^*}-\frac{1}{g_n (R_n^{*})^2}f'\left(\frac{\ell_n^{(s)}}{t_n^* R_n^*}\right)=0.
\end{equation*}
Combining the above with the optimal conditions,  $\widehat{\lambda}^*=-\frac{1}{g_n}g\left(\frac{\ell_n^{(s)}}{t_n^* R_n^*}\right)$ and $t_n^*=T-\frac{\ell_n^{(s)}\left(e^{\epsilon R_n^*}-e^{\epsilon}\right)}{f_n}-\frac{\ell_n^{(s)}}{s_n}$ according to \eqref{Eq:LosslessKKT2} and \eqref{Eq:LosslessKKT3}, produces the desired result.

\subsection{Proof of Lemma~\ref{Lem:Losslessmonotone}}\label{App:Losslessmonotone}
The first derivative of function $z({R_n})$ is
\begin{align*}
\frac{\partial z(R_n)}{\partial R_n}&
\!=\!\frac{N_0 \ln2}{B R_n^2 g_n}\left[\frac{2}{R_n}+\frac{\ln2[R_n d(R_n)]'}{B [R_n d(R_n)]^2}\left(1\!-\!\frac{R_n d'({R_n})}{d(R_n)}\right)\right]\\
&~~~ \times\!e^{\frac{\ln2}{B R_n d(R_n)}}\!+\!\left[q_n^{(c)}\!-\!\frac{1}{g_n f_n}g\left(\frac{1}{d(R_n)R_n}\right)\right]\epsilon^2 e^{\epsilon R_n}\!.
\end{align*}
First, it has $[R_n d({R_n})]^{'}\!=\!d({R_n})\!-\!R_n d'({R_n})$. Hence, $[R_n d({R_n})]'>0$ when $d({R_n})>R_n d'({R_n})$ and $[R_n d({R_n})]'<0$ when $d({R_n})<R_n d'({R_n})$. Combining these together leads to: 
\begin{equation*}
\frac{\ln2[R_n d(R_n)]'}{B [R_n d(R_n)]^2}\left(1-\frac{R_n d'({R_n})}{d(R_n)}\right)>0, ~~\text{for}~ R_n\in[1, R_{\rm{max}}].
\end{equation*} Further, the first derivative of function $g(x)$ is 
\begin{equation*}
g'(x)=-x f''(x)=-\frac{N_0\ln2^2}{B^2}2^{\frac{x}{B}}<0, ~~\text{for}~x>0.
\end{equation*}
It means that $g(x)$ is monotone-decreasing with $x$ and thus $g(x)\le g(x)|_{x=0}=0$ for  $x>0$. Since $\frac{1}{d(R_n)R_n}>0$, it has  $\left[q_n^{(c)}-\frac{1}{g_n f_n}g\left(\frac{1}{d(R_n)R_n}\right)\right]>0$ for $x>0$. Combining the above considerations leads to  $\frac{\partial z(R_n)}{\partial R_n}>0$ for $R_n\in[1, R_{\rm{max}}]$, thus completing
 the proof.

\subsection{Proof of Corollary~\ref{Cor:CompressionRatio}}\label{App:CompressionRatio}
\par Define a function $\phi(R_n)$ as $\phi(R_n)=q_n^{(c)} g_n$. According to \eqref{Eq:LosslessR}, it has
\begin{equation*}
\phi(R_n)= \frac{1}{f_n}g\left(\frac{1}{d(R_n)R_n}\right)+\frac{1}{R_n^2 \epsilon e^{\epsilon R_n}}f'\left(\frac{1}{d(R_n)R_n}\right).
\end{equation*}
The derivation of $\phi(R_n)$ is
\begin{equation*}
\begin{aligned}
\frac{\partial \phi(R_n)}{\partial R_n}
&=-\left[\frac{2+\epsilon}{R_n \epsilon e^{\epsilon R_n}}+\frac{\ln2[R_n d(R_n)]'}{B [R_n d(R_n)]^2}\left(1-\frac{R_n \epsilon e^{\epsilon R_n}}{f_n d(R_n)}\right)\right]\\
&\quad\quad\quad \times\frac{N_0 \ln2}{B R_n^2 \epsilon e^{\epsilon R_n}}e^{\frac{\ln2}{B R_n d(R_n)}}.
\end{aligned}
\end{equation*}
Since $\frac{\ln2[R_n d(R_n)]'}{B [R_n d(R_n)]^2}\left(1-\frac{R_n \epsilon e^{\epsilon R_n}}{f_n d(R_n)}\right)$ is always positive as proved in Appendix~\ref{App:Losslessmonotone}, the function $\phi(R_n)$ is monotone-decreasing for $R_n\!\in\![1, R_{\rm{max}}]$. Given $q_1^{(c)}\!>\!q_2^{(c)}$, it can be derived that $R_1\!<\!R_2$. Using a similar approach, it can be proved that  $R_1\!<\!R_2$ when $g_1\!>\!g_2$,  thus completing  the proof.

\subsection{Proof of Lemma~\ref{Lem:Lossymonotone}}\label{App:Lossymonotone}
The first derivative of $\check{z}({r_n})$ is
\begin{equation*}
\begin{aligned}
&\frac{N_0 \ln2}{B g_n r_n^2}\left[\frac{2}{r_n}+\frac{\ln2[r_n \check{d}(r_n)]'}{B [r_n \check{d}(r_n)]^2}\left(1-\frac{r_n \check{d}'({r_n})}{\check{d}(r_n)}\right)\right]e^{\frac{\ln2}{B r_n \check{d}(r_n)}}+\\
&\quad (6+4\epsilon r_n^2)r_n q_n^{(c)} \epsilon e^{\epsilon r_n^2}-g\left(\frac{1}{\check{d}(r_n)r_n}\right)\frac{(6+4\epsilon r_n^2)\epsilon r_n}{g_n f_n}e^{\epsilon r_n^2}.
\end{aligned}
\end{equation*}
The derivative of function $r_n \check{d}({r_n})$ is $\check{d}({r_n})-r_n \check{d}'({r_n})$. Since $[r_n \check{d}({r_n})]'>0$ when $\check{d}({r_n})>r_n \check{d}'({r_n})$ and $[r_n \check{d}({r_n})]'<0$ when $\check{d}({r_n})<r_n \check{d}'({r_n})$, the item $\frac{\ln2[r_n \check{d}(r_n)]'}{B [r_n \check{d}(r_n)]^2}\left(1-\frac{r_n \check{d}'({r_n})}{\check{d}(r_n)}\right)$ is always positive. Moreover, $g(x)<0$ when $x>0$ according to Appendix~\ref{App:Losslessmonotone}. Since $\frac{1}{\check{d}(r_n)r_n}>0$, the item $\left[(6+4\epsilon r_n^2)r_n q_n^{(c)} \epsilon e^{\epsilon r_n^2}-g\left(\frac{1}{\check{d}(r_n)r_n}\right)\frac{(6+4\epsilon r_n^2)\epsilon r_n}{g_n f_n}e^{\epsilon r_n^2}\right]$ is always positive. Hence, $\frac{\partial \check{Z}(r_n)}{\partial r_n}>0$ for $r_n\in[1, r_{\rm{max}}]$, thus completing the proof.

\bibliographystyle{ieeetr}

\end{document}